\documentclass[preprint,showpacs,preprintnumbers,amsmath,amssymb,superscriptaddress,nofootinbib]{revtex4}
\usepackage{graphicx,color}
\usepackage{amsmath,amssymb}
\usepackage{url}
\usepackage{epstopdf}
\newcommand{\hs}{\hspace*{0.5cm}}

\newcommand{\be}{\begin{equation}}
\newcommand{\ee}{\end{equation}}
\newcommand{\bea}{\begin{eqnarray}}
\newcommand{\eea}{\end{eqnarray}}
\newcommand{\nn}{\nonumber}
\newcommand{\crn}{\nonumber \\}

\newcommand{\la}{\lambda}

\newcommand{\ga}{\gamma}

\newcommand{\om}{\omega}

\newcommand{\bc}{\begin{center}}
\newcommand{\ec}{\end{center}}
\newcommand{\Ga}{\Gamma}

\newcommand{\La}{\Lambda}

\newcommand{\ps}{\psi}

\newcommand {\ba}{\begin{array}}
\newcommand {\ea}{\end{array}}
\newcommand{\ben}{\begin{enumerate}}
\newcommand{\een}{\end{enumerate}}

\usepackage{bm}
\usepackage{dcolumn}

\begin{document}
\title{CP violations in a predictive $A_4$ symmetry model}
\author{T. Phong Nguyen}\email{thanhphong@ctu.edu.vn}
\affiliation{Department of Physics, Can Tho University,
	3/2 Street, Can Tho, Vietnam}%%%%%%%%%%%
\author{L.~T.~Hue }\email{lethohue@duytan.edu.vn}
\affiliation{Institute of Research and Development, Duy Tan University,  Da Nang 550000, Vietnam}

\author{D. T. Si}\email{dangtrungsi@duytan.edu.vn}
\affiliation{Institute of Research and Development, Duy Tan University,  Da Nang 550000, Vietnam}
\author{T. T. Thuc}\email{truongtrongthuck17@gmail.com}
\affiliation{Department of Education and Training of Ca Mau, 70 Phan Dinh Phung, Vietnam}
%%%%%%%%%
\begin{abstract}
We will investigate numerically a seesaw model with $A_4$ flavor symmetry to find allowed regions satisfying the current experimental neutrino oscillation data, then  use them to predict physical consequences.   Namely, the lightest active neutrino mass  is of
 the  order of $\mathcal{O}(10^{-2})$ eV.  The effective neutrino mass $|\langle m\rangle|$ associated with neutrinoless double beta decay is in the range $[0.002 \;\mathrm{eV},0.038\;\mathrm{eV}]$  and  $[0.048\;\mathrm{eV},0.058\;\mathrm{eV}]$, corresponding to the normal and the inverted hierarchy schemes, respectively. Other relations among relevant physical quantities are shown, so that they can be determined if some of them are confirmed experimentally.   The recent data of the baryon asymmetry of the Universe ($\eta_B$) can be explained via   leptogenesis caused by  the  effect of the renormalization group evolution on the Dirac Yukawa couplings, provided the right-handed neutrino mass scale $M_0$   ranges from $\mathcal{O}(10^8)$  GeV to $\mathcal{O}(10^{12})$ GeV for $\tan\beta =3$. This allowed $M_0$  range is different from the scale of  $\mathcal{O}(10^{13})$  GeV for other effects that also generate a consistent $\eta_B$  from leptogenesis.  The branching ratio of the decay $ \mu \rightarrow\,e\gamma$  may reach future experimental sensitivity for very light values of $M_0$. Hence, it will be  inconsistent with the $M_0$ range predicted from the $\eta_B$ data whenever  this decay is detected experimentally.
%\keywords{$A_4$ model; CP violation; Leptogenesis}
\end{abstract}
%\pacs{11.30.Hv, 13.35.Hb, 14.60.Pq, 14.60.St}
\maketitle
%%%%%%%%%%%%%%%%%%%
\allowdisplaybreaks
\section{Introduction}
The  experimental data for neutrino oscillation definitely affirmed that neutrinos are massive  and they are mixing. Based on neutrino experimental data, in 2002, P.  F. Harrison et al.~\cite{Harrison:2002er,Harrison:2002kp,Harrison:2002et,Harrison:2003aw}  proposed the structure of neutrino mixing matrix  named tri-bimaximal (TB). According to this structure, the reactor mixing angle, $\theta_{13}$, is zero and the Dirac CP-violating phase has no meaning. Subsequently,  there was  a lot of effort to build simple models leading to the TB mixing pattern of leptons. An interesting way seems to be the use of some discrete non-Abelian flavor groups added to the gauge group of the Standard Model (SM).  There is a series of such models based on the symmetry groups $A_4$ \cite{Ma:2001dn,Babu:2002dz,Altarelli:2005yp,Altarelli:2005yx,Bazzocchi:2007na,Brahmachari:2008fn,Adhikary:2008au},  $T'$~\cite{Feruglio:2007uu,Chen:2007afa,Frampton:2007et,Frampton:2009fw},  and $S_4$ \cite{Pakvasa:1978tx,Brown:1984mq,Lee:1994qx,Ma:2005pd}.
  These models are usually realized at some high-energy scale $\Lambda$ and the groups are spontaneously broken due to a set of scalar multiplets. On the other hand,  the most up-to-date data  from neutrino oscillation experiments shows that  the reactor mixing angle is relatively large, $\theta_{13} \sim 8^{\textrm{o}}$~\cite{Tanabashi:2018oca}. As a result, the models mentioned  have been improved in order to generate a non-zero value of  $\theta_{13}$ as well as leptogenesis;  see, for example  the models with  $A_4$ symmetry given in Refs.~\cite{Altarelli:2005yx,Altarelli:2012ss,Ma:2012xp,Chen:2012st,Ahn:2013mva},  where higher-order corrections to fermion mass matrices were considered.
  However, according to these works, just the inclusion of higher-order corrections would not produce such a large value of $\theta_{13}$ consistent with experiment.  Improved models with modular $A_4$ symmetry groups have also been constructed recently to explain the neutrino oscillation data~\cite{Novichkov:2018yse,Ding:2019zxk}.  On the other side, several models were built by adding new sources of $A_4$ breaking at the leading orders  into the original $A_4$ models~\cite{Altarelli:2005yx},  so that they can  esuccessfully xplain both  experimental values of  $\theta_{13}$ and  leptogenesis; see, for example, Refs. \cite{Adhikary:2008au,Karmakar:2015jza, Chen:2012st, Morisi:2013qna, Karmakar:2014dva},  and a list of other models reviewed  in Ref.~\cite{Barry:2010zk}.   In particular, a soft breaking $A_4$ term was introduced in Ref.~\cite{Adhikary:2008au}, three  singlet flavons were used in Refs. \cite{Karmakar:2015jza, Chen:2012st}, and two singlet flavons $\xi, \xi'$ transform as $1, 1'$ of the $A_4$ in Refs.~\cite{Chen:2012st, Morisi:2013qna, Karmakar:2014dva} in order to accommodate with the present neutrino data. However, leptogenesis was not studied in Ref.  \cite{Morisi:2013qna}, while in Ref.~\cite{Karmakar:2014dva}, to explain conventional leptogenesis  the authors considered the contribution of the next-to-leading order (NLO) corrections to the right-handed neutrino (RHN) mass matrix  in the suppersymmetry framework. Namely, two new NLO terms corresponding to two new independent parameters were introduced by hand, then their allowed values were investigated to guarantee successful leptogenesis, leading to a prediction that the RHN mass scale is around $\mathcal{O}(10^{13})$ GeV. But these terms will not survive in other models where new charge assignments of discrete symmetries  are chosen to cancel them. In addition, it seems that this approach still needs more independent parameters than an alternative presented in Ref.~\cite{Adhikary:2008au}, where a single softly broken $A_4$ term was added into the original model to successfully solve both neutrino data and leptogenesis, leading to a prediction of  the RHN mass scale  of $O(10^{13})$ GeV.  The model mentioned in  Ref.~\cite{Adhikary:2008au} is the simplest extension of the original one discussed in Ref.~\cite{Altarelli:2005yx}.

In this work we will study another simple approach, based on the model introduced in Ref. \cite{Karmakar:2014dva}, but  only effects of  renormalization group (RG) evolution of the Dirac Yukawa coupling matrix will be included  to study flavored leptogenesis. Because only two flavon singlets are added into the model, and the NLO terms like those mentioned in Ref.~\cite{Karmakar:2014dva} are excluded by the total symmetry, only one new term appears in the model, therefore it is also as simple as the model given in Ref.~\cite{Adhikary:2008au}. We believe that the effect caused by just the RG is as important as the effects  arising from the NLO terms mentioned in Ref.~\cite{Karmakar:2014dva},  where successful leptogenegis  requires a large RHN mass scale of around $\mathcal{O}(10^{13})$  GeV. Also, the same RHN mass scale is needed for successful leptogenesis in the model constructed in Ref.~\cite{Adhikary:2008au}. This scale is only three orders less than the perturbative limit of the seesaw (SS) model \cite{Minkowski:1977sc,Mohapatra:1979ia,GellMann:1980vs,Yanagida:1979as,Schechter:1980gr}  $\sqrt{4\pi}\times174/10^{-12}\sim \mathcal{O}(10^{16})$. Our work will find an interesting answer for the question of whether the RG effects need a lower RHN mass scale to explain leptogenesis, or whether they have the same order of $10^{13}$ GeV predicted previously.  Anyway, we can discuss which effects are dominant or whether there any properties to distinguish these two effects if they appear simultaneously in the same RHN mass scale. The correlation of the two effects will also be very interesting, but we will leave this for further study.

Besides generating a tiny neutrino mass, the seesaw model has another physics consequence called leptogenesis for the generation of the observed baryon asymmetry of the Universe (BAU) by the charge/parity (CP) asymmetric decay of heavy RHNs \cite{Fukugita:1986hr, Giudice:2003jh,Buchmuller:2004nz}.  If the BAU was generated by leptogenesis, then CP-violation in the lepton sector must exist.
For Majorana neutrinos, there are one Dirac and two Majorana CP violating phases. One of the phases (or a combination of them) in principle can be measured by neutrinoless double beta ($0\nu2\beta$) decay \cite{Bilenky:2001rz, Pascoli:2001by,Pascoli:2002qm,Petcov:2004wz} experiments. Also, the TB mixing structure forbids low-energy CP violation in neutrino oscillation, due to $U_{e3}=0$, and also forbids high-energy CP violation in leptogenesis. Therefore, any observations  of leptonic CP violation, for instance in $0\nu2\beta$ decay, can strengthen our believe in leptogenesis by demonstrating that CP is not a  lepton symmetry.

In this work we consider an expansion of the SM by the seesaw realization of an $A_4$ discrete symmetric model and its phenomena. Apart from two SM scalar doublets taking responsibility for spontaneously breaking of $A_4$ and the SM gauge groups, this model contains additional $SU(2)_L$ scalar singlets, namely two singlets $\xi', \xi''$ transform as $1', 1''$ and two triplets of the $A_4$. If the RHN mass matrix's components resulting from the contributions of the vacuum expectation values (VEVs) of two scalar singlets (of both $SU(2)_L$ and $A_4$) are exactly the same, then the model generates the TB pattern of lepton mixing matrix and hence leptogenesis does not work. We therefore study the case where those components are independent, and we find the allowed regions of the parameter space of the model that satisfy the low-energy data and  the recent  BAU data through flavored leptogenesis that arise from the RG effects at a high scale of RHN masses. At low-energy, although our model inherits similar properties in the lepton sector to some previous works \cite{Karmakar:2014dva}, where some of the unknown parameters were fixed to determine the allowed regions, in this work we will scan the whole parameter space to collect all possible allowed regions satisfying the recent neutrino oscillation data. Based on this,  we give interesting physical consequences of $|\langle m\rangle|$  and lepton flavor violating (LFV) decays.  We will also determine the RHN mass scale at high energy that successfully explains the BAU data originating from just the RG effect. The allowed range of RHN mass scale will be used to compare with those concerned previously, which come from other sources of soft breaking $A_4$ or NLO terms that are forbidden in our model.

This work is organized as follows. In Sect. \ref{sec_a4model} we summarize all the ingredients for constructing the $A_4$ model with the seesaw mechanism, focusing on the Higgs and lepton sectors. After that,  we present, step by step, our approach to numerically investigating the parameter space of the model to guarantee that  all allowed regions satisfying the recent neutrino oscillation data are pointed out. Following this, we continue predicting some consequences related to the low-energy phenomena of the lepton sector. Section \ref{sec_ Leptogenesis} is devoted to studying the leptogenesis originating purely from the RG effects, where the allowed range of the RHN mass scale that satisfies the BAU data will be determined. Section \ref{sec_cLFV} will pay attention to the LFV decay of charged leptons, and will show that these decays can be considered as another indirect channel to estimate the RHN mass scale.  Important conclusions from our work are given in the last section, section V. In addition, there are three appendices to add more detailed discussions on the $A_4$ rules, the Higgs potential, and analytic formulas for one-loop contributions to the LFV decays in the unitary gauge.

%%%%%%%%%%%%%%%%%%%
\section{ \label{sec_a4model} The $A_4$ symmetry model with seesaw mechanism}
The non-Abelian $A_4$ is a group of even permutations of four objects and has $4!/2=12$ elements.  All the properties of this group needed for model construction were given in Ref. \cite{Altarelli:2005yx}. This paper will work in the $A_4$ basis introduced by  G. Altarelli and F. Feruglio, as reviewed in Appendix~\ref{A4rules}.
In this work we promote the $A_4$ proposed in Refs. \cite{Adhikary:2008au,Karmakar:2014dva} with two Higgs singlets to accompany the seesaw mechanism. The model contains several $SU(2)_L\otimes U(1)_Y$ Higgs singlets, where two of them ($\xi',\ \xi''$) are $A_4$ singlets, while the remaining ($\phi_S,\ \phi_T$) are triplets. The SM lepton doublets are assigned to be three components of one $A_4$ triplet, while three right-handed charged leptons $e_R,\mu_R,\tau_R$ are assumed to transform as three different singlets $1, 1'', 1'$, respectively. The standard Higgs doublets $h_u$ and $h_d$ remain invariant under $A_4$. The particle content for leptons and scalars, their VEVs, and the symmetry groups considered in the model are shown in Table \ref{particle content}.  Two more discrete symmetries, $Z_{3}$ and $Z_4$, are included in order to get minimal and necessary Yukawa couplings.
%30/Oct/19, change Z4 charge of eR,muR,tauR and phiT from 1 to (-1)
\begin{table}[ht]
	\caption{List of fermion and scalar fields, where $\overline{\psi^l}=(\overline{\nu_{La}},\;\overline{e_{La}})^T$ ($a=1,2,3$) and $\omega= e^{2i\pi/3}$. }
		{\begin{tabular}{lcccccc}\hline\hline
		Lepton & $SU(2)_L$   &   $U(1)_Y$ &  $A_4$  & $Z_3$ & $Z_4$ & \\ \hline
		$\overline{\psi^l}$ & $\ 2^*$ & 1&$\underline{3^*}$  & 1 & 1\\
		$e_R$ &    1    &  $-2$  &\underline{1 }  &  1 & $-1$\\
		$\mu_R$ &    1    & $-2$&    $\underline{1}'$   &1  & $-1$ \\
		$\tau_R$ &    1    & $-2$&    $\underline{1}''$   & 1 & $-1$ \\
	$N_{R}$&    1    & 0&    $\underline{3}$   &  $\om$ & $-i$ \\
		\hline
		Scalar   &         &   &   &  &  &  VEV \\
		\hline
		$h_u$    &    2    &$-1$ &   \underline{ 1}          &$\om^2$    & $i$ &  $\langle h_u\rangle=v_u$                           \\
		$h_d$     &    2    & 1 &    \underline{1}          &    1         &  1  &  $\langle h_d\rangle=v_d$                           \\
		$\phi_S$  &    1    & 0 &    \underline{3}          &  $\om$ &  $-1$ &  $\langle\phi_S\rangle=(v_S,v_S,v_S)$ \\
		$\phi_T$  &    1    & 0 &    \underline{3}          & 1            &   $-1$ &  $\langle\phi_T\rangle=(v_T,0,0)$                   \\
		$\xi'$     &    1   &0 &    $\underline{1}'$  & $\om$  &  $-1$ &  $\langle\xi'\rangle=u'$                             \\
		$\xi''$    &    1   &0 &    $\underline{1}''$ & $\om$  & $-1$  &  $\langle\xi^{''}\rangle=u''$
		\\
		\hline \hline
	\end{tabular}
\label{particle content}}
\end{table}

The Lagrangian for the lepton sector which is invariant under all the symmetries given in Table \ref{particle content} is
\bea
\label{lagrangian}
-{\cal L}&=& \frac{y_e}{\La}(\phi_T\bar{\psi}_L^l)e_Rh_d+\frac{y_\mu}{\La}(\phi_T\bar{\psi}_L^l)''\mu_Rh_d
+\frac{y_\tau}{\La}(\phi_T\bar{\psi}_L^l)'\tau_Rh_d+p\bar{\psi}_L^lN_Rh_u\crn
&& +x_A'\xi'(\bar{N}_L^cN_R)''+x_A''\xi''(\bar{N}_L^cN_R)'+x_B(\phi_S\bar{N}_L^cN_R)+{\rm H.c.}, \label{Lpartlep}
\eea
where $N^c_L\equiv C(\bar{N}_R)^\textrm{T}$ and $\Lambda$ is the cut-off scale of the model. It can be seen that the NLO term like $(\bar{\psi}_L^lN_R\phi_Th_u/\Lambda)$  mentioned in Ref.~\cite{Karmakar:2014dva} does not respect the $Z_4$ symmetry, hence this term vanishes in our model.
After spontaneous symmetry breaking, the charged lepton mass matrix comes out diagonally with
$m_e=\frac{y_e v_T v_d}{\La}$, $m_\mu=\frac{y_\mu v_T v_d}{\La}$, and $m_\tau=\frac{y_\tau v_T v_d}{\La}$. The couplings $y_e, y_\mu$, and $y_\tau$ are naturally the same order of magnitude. In order to produce the mass hierarchy of charged leptons, we make use of an additional spontaneously broken $U(1)_{FN}$ flavor \cite{Froggatt:1978nt}. We introduce a singlet $\theta$ carrying $U(1)_{FN}$ charge $-1$ and neutral under all other symmetries. Its VEV, $\langle \theta\rangle / \Lambda < 1$, breaks $U(1)_{FN}$ and provides expansion parameters for charged lepton masses. We also assign $U(1)_{FN}$ charges $(2n, n)$ to fields $(e_R, \mu_R)$. All other lepton fields are assigned to be neutral under this symmetry. In this way, $y_\tau:y_\mu:y_e = 1: (\langle \theta\rangle / \Lambda)^n: (\langle \theta\rangle / \Lambda)^{2n}$, and the charged lepton mass hierarchy can be produced by choosing $(\langle \theta\rangle / \Lambda)^n \simeq \lambda^2$, where $\lambda \simeq 0.225$ is the Wolfenstein parameter.

Regarding the quark sector,  under the symmetry $SU(3)_C\otimes SU(2)_L\otimes U(1)_Y\otimes A_4\otimes Z_3\otimes Z_4$,  they can be assigned as follows:  $Q_{iL}=(u_i,\;d_i)_L^\textrm{T}\sim (3,2,1/3,\underline{1},1,1)$, $u_{iR}\sim (3,1,4/3,\underline{1},w,-i)$, and   $d_{iR}\sim (3,1,-2/3,\underline{1},w^2,i)$. Accordingly, the Yukawa Lagrangian of the quarks has the same form as given in the SM, namely
\begin{align}
\label{eq_LYq}
\mathcal{L}^Y_{q}&=-Y^u_{ij}\overline{Q_{iL}}h_uu_{jR} -Y^d_{ij}\overline{Q_{iL}}\tilde{h}_ud_{jR} +\mathrm{h.c.},
\end{align}
where $\tilde{h}_u=i\sigma_2 h^*_u$. Although the phenomenology of the quark will not be considered in this work, the Yukawa couplings of the top quark with $h_u$ given in Eq.~\eqref{eq_LYq} will give a top quark mass  $m_t\simeq Y^q_{33}v_u$, which requires large $v_u$ to generate the well-known top quark mass while  $Y^q_{33}$  satisfies the perturbative limit.  This also implies a large $v_u/v_d$, which  will be chosen for numerical investigation.

Before continuing to the lepton sector, we note that the VEV structure of the scalar fields assumed in Table~\ref{particle content} are realistic; see a detailed discussion on the Higgs potential in Appendix \ref{HiggsPotential}.  We have also shown that the model contains an SM-like Higgs boson found experimentally by LHC \cite{Aad:2012tfa, Chatrchyan:2012xdj}.

For the charged Higgs boson, in the basis $(H^\pm_{u},\;H^\pm_{d})^T$ the squared mass matrix is
\bea
M^{2}_{\rm charged}=\la_4\times \left(\begin{array}{cc}
	v_d^2  &  v_dv_u  \\
	v_dv_u  & v_u^2  \\
\end{array}\right). \eea
It gives two pairs of mass eigenstate denoted as  physical Higgs bosons $\varphi^\pm$ and massless states $G^\pm$ which are Goldstone bosons eaten by gauge bosons $W^\pm$. The masses and relations between the original and mass base of the charged Higg components are as follows,
\bea
&& m^2_{G^\pm}=0, \hs G^\pm=s_\beta H_u^\pm-c_\beta H_d^\pm, \crn
&& m^2_{\varphi^\pm}=\la_4 v^2, \hs \varphi^\pm=c_\beta H_u^\pm+s_\beta H_d^\pm, \label{cHiggs}
\eea
where $s_{\beta}\equiv\sin\beta$, $c_{\beta}\equiv\cos\beta$, and $\beta$ is a mixing angle defined by
\be
t_{\beta}\equiv\tan\beta= \frac{v_u}{v_d},\label{tbeta}
\ee
which is similar to the case of the ratio defined by the two VEVs  in the minimal supersymmetric Standard Model (MSSM). The charged Higgs boson and the parameter $\beta$ play very important roles for generating leptogenesis in our model. We emphasize that these two ingredients are independent from all Higgs self couplings of the $SU(2)_L$ Higgs singlets.

To identify the gauge bosons with those in the SM, we start from the  covariant derivative for  local $SU(2)_L\otimes U(1)_Y$ symmetry. It is defined  as
\be D_{\mu} = \partial_{\mu} - igT^aW^a_{\mu}-i\frac{g'}{2}B_{\mu}Y, \label{derivative}\ee
which is the same as in the SM. Here, $T^a$ ($a=1,2,3$) are the generators of the $SU(2)_L$ symmetry. $T^a=\frac{\sigma^a}{2}$ for doublets, and $T^a=0$ for singlets.

The kinetic terms of all Higgses are
\bea  \mathcal{L}^H_{\rm{kin}}&=& \left(D_{\mu}h_u\right)^{\dag}\left(D^{\mu}h_u\right) + \left(D_{\mu}h_d\right)^{\dag}\left(D^{\mu}h_d\right)\crn
&&+ \left[(\partial_{\mu}\phi_T)^{\dag}\partial^{\mu}\phi_T\right]_{\underline{1}}+ \left[(\partial_{\mu}\phi_S)^{\dag}\partial^{\mu}\phi_S\right]_{\underline{1}} + \partial_{\mu}\xi'\partial^{\mu}\xi' + \partial_{\mu}\xi''\partial^{\mu}\xi''.\label{Hkinetic}\eea
From Table \ref{particle content}, where all neutral Higgs singlets have zero $U(1)_Y$ charges,   we can see that all $SU(2)_L$ singlets do not couple with gauge bosons. The mass term of the gauge bosons is
\be  \mathcal{L}^{\mathrm{gauge}}_{m}= \frac{g^2(v_u^2+v_d^2)}{2} W^{+\mu}W^-_{\mu}+\frac{g^2(v_u^2+v_d^2)}{4}\left(W_3-t_W B\right)^{\mu}\left(W_3-t_WB\right)_{\mu}, \label{Gmass}\ee
where $W^{\pm}\equiv (W^1\mp i W^2)/\sqrt{2}$ and  $t_W\equiv g'/g$. Matching with the mass of the  SM  gauge boson $W^\pm$ we obtain the same relation shown in two Higgs doublet models (2HDMs),
\be  v^2=v_u^2+v_d^2=174^2\mathrm{ GeV}^2, \hs t_W=\frac{s_W}{c_W}, \label{SMmatching}\ee
where $s_W^2=0.231$. It is easy to show that the second term in Eq.~\eqref{Gmass} implies the presence of  the photon and  neutral $Z$ boson defined in the SM.

Regarding the lepton, the neutrino sector gives rise to the following Dirac and Majorana neutrino mass matrices:
\bea
\label{Majoranamass1}
m_D&=& p v_u{\left(\begin{array}{ccc}
		1  & 0  &  0\\
		0  &   1  &  0 \\
		0  &   0  & 1 \end{array}\right)}=v_uY_\nu, \quad Y_\nu = p\times \textbf{1},
\eea
\bea
\label{Majoranamass2}
M_R = {\left(\begin{array}{ccc}
		\frac{2X}{3} & \tilde{Z}-\frac{X}{3} & \tilde{Y}-\frac{X}{3}\\
		\tilde{Z}-\frac{X}{3}& \tilde{Y}+\frac{2X}{3} & -\frac{X}{3} \\
		\tilde{Y}-\frac{X}{3} & -\frac{X}{3} & \tilde{Z}+\frac{2X}{3}\end{array}\right)}=
M_0{\left(\begin{array}{ccc}
		1 & \tilde{\kappa} -\frac{1}{2} &  \tilde{\rho} -\frac{1}{2}\\
		\tilde{\kappa} -\frac{1}{2} & \tilde{\rho} +1 & -\frac{1}{2} \\
		\tilde{\rho} -\frac{1}{2} & -\frac{1}{2} & \tilde{\kappa} +1\end{array}\right)}  ,
\eea
where $X=2x_B v_S,\ \tilde{Y}=2x_A'u',\ \tilde{Z}=2x_A''u''$, and  $M_0=2X/3$ is the scale of the RHN mass, $\tilde{\kappa} =\tilde{Z}/M_0,\ \tilde{\rho} = \tilde{Y}/M_0$. We assume that $M_0 $ is real and positive. Hereafter, complex parameters are distinguished by tildes. Then, the active neutrino mass matrix is then obtained by the seesaw  formula \cite{Minkowski:1977sc,Mohapatra:1979ia,GellMann:1980vs,Yanagida:1979as,Schechter:1980gr}:
\be
\label{active mass}
m_\nu = - v_u^2 Y_\nu^\textrm{T} M_R^{-1} Y_\nu.
\ee
This matrix is used to determine the active neutrino masses $m_{1,2,3}$ and neutrino mixing matrix $U_{\nu}$, namely
\begin{equation}\label{eq_lightactivenu}
U_\nu^\textrm{T} m_\nu U_\nu = {\rm diag}\left(m_1,~m_2,~m_3 \right) \equiv m_\nu^d,
\end{equation}
where $m_{1,2,3}$ is positive real, and the lepton mixing matrix $U_{\rm PMNS} = U_\nu$ since the charged lepton mass matrix is diagonal in our case.  For later convenience, at first we diagonalize the right-handed neutrino mass matrix $M_R$ based on the nearly TB forms discussed previously. In particular, if $\tilde{\rho} = \tilde{\kappa}$ then $M_R$ is exactly diagonalized by the well-known TB structured matrix, namely
\begin{align}
\label{eq_UTB}
U_{\mathrm{TB}}= \left(
\begin{array}{ccc}
\sqrt{\frac{2}{3}} & \frac{1}{\sqrt{3}} & 0 \\
-\frac{1}{\sqrt{6}} & \frac{1}{\sqrt{3}} & -\frac{1}{\sqrt{2}} \\
-\frac{1}{\sqrt{6}} & \frac{1}{\sqrt{3}} & \frac{1}{\sqrt{2}} \\
\end{array}
\right).
\end{align}
%Here we set a tinny shift between the two parameter, namely $\tilde{\rho} = \tilde{\kappa}(1+\epsilon)$, $\epsilon$ is a small real parameter. Note that, this choice implies that ${\rm arg}(\tilde{\rho})={\rm arg}(\tilde{\kappa})=\phi$.
Hence, the non-zero $s_{13}$ may arise from the deviation of these two parameters. Furthermore, $s_{13}$ is found experimentally to be rather smaller than 1, which suggests that the deviation should be small. Hence, in this work we  adopt that $\tilde{\rho} = \tilde{\kappa}(1+\epsilon)$, where $\epsilon$ is a  complex parameter satisfying $|\epsilon| <1$, so that  it will be used as a reliable perturbative parameter in the  next approximate calculations. Then the matrix $M_R$ is rewritten in a new form as
\bea
\label{Majoranamass2}
M_R &= M_0\left( \begin{array}{*{20}{c}}
1&\tilde{\kappa }- \frac{1}{2}&\tilde{\kappa}(1+\epsilon) - \frac{1}{2}\\
\tilde{\kappa} - \frac{1}{2}&\tilde{\kappa}(1+\epsilon)  + 1&- \frac{1}{2}\\
\tilde{\kappa}(1+\epsilon)- \frac{1}{2}&- \frac{1}{2}&\tilde{\kappa}  + 1
\end{array} \right).
\eea
This matrix is diagonalized by a unitary matrix $U_R$ defined as follows
\bea
\label{D_MR}
M_R^d &=& U_R^\textrm{T}M_RU_R = {\rm diag}\left(M_1,~M_2,~M_3\right),
\eea
where
\bea
\label{MR_eigenvalues}
M_1 &=& M_0\left|\frac{1}{2}(3-2\sqrt{\tilde{\kappa}^2\epsilon^2+\tilde{\kappa}^2\epsilon+\tilde{\kappa}^2})\right|
     =M_0\left|\frac{1}{2}(3-2\tilde{\kappa}\sqrt{1+\epsilon+\epsilon^2})\right|,\nn\\
M_2 &=&M_0 \left|\tilde{\kappa}(2+\epsilon) \right|,\\
M_3 &=& M_0\left|\frac{1}{2}(3+2\sqrt{\tilde{\kappa}^2\epsilon^2+\tilde{\kappa}^2\epsilon+\tilde{\kappa}^2})\right|
     =M_0\left|\frac{1}{2}(3+2\tilde{\kappa}\sqrt{1+\epsilon+\epsilon^2})\right|\nn
\eea
 are masses of physical RHNs, and the matrix $U_R$ is determined  through two steps, where the first relates to  $U_{\mathrm{TB}}$, while the second is a product of a unitary matrix $U_1$ depending on $\epsilon$ and  a phase matrix $U_P$. The precise form is
\bea
U_R &=& U_{\rm TB}U_1U_P,\nn\\
U_1 &=& \left( \begin{array}{ccc}
c_{\theta} &0 & s_{\theta} e^{i\zeta}\\
 0 &  1&  0\\
- s_{\theta} e^{-i\zeta} & 0 & c_{\theta}
\end{array} \right),~ U_P = \left( \begin{array}{ccc}
e^{-i\varphi_1/2} & 0 & 0\\
 0 &  e^{-i\varphi_2/2} &  0\\
 0 & 0 & e^{-i\varphi_3/2}
\end{array} \right), \label{U_R matrix} \\
\varphi_1 &=&\arg{(3-2\tilde{\kappa}\sqrt{1+\epsilon+\epsilon^2})},~\varphi_2 ~ =~\arg{\tilde{\kappa}(2+\epsilon)}=\phi,\nn\\
\varphi_3 &=& \arg{(3+2\tilde{\kappa}\sqrt{1+\epsilon+\epsilon^2})},\label{MR_phases}
\eea
where $s_{\theta}$ is positive and $\zeta$ is real. We note that the consideration that $M_{1,2,3}$ are physical RHNs is consistent in the seesaw limit, as we will point out later.

Because of the diagonal form of the matrix $Y$ in the light neutrino mass matrix $m_\nu$ given in Eq. (\ref{active mass}), it is diagonalized exactly using $U_{R}$ through the following intermediate transformation:
\bea
m_{\nu}&=& v_u ^2 Y_\nu^\textrm{T} \left(U_R^\ast M_R^d U_R^\dag \right)^{-1}Y_\nu\nn\\
       &=& U_R {\rm diag}(m_1,~m_2,~m_3)U_R^\textrm{T} ~\equiv~  U_\nu^\ast m_\nu^d U_\nu^\dag, \label{eq_mnu1}
\eea
 where $m_\nu^d $ was defined previously in Eq.~\eqref{eq_lightactivenu}, and the active neutrino masses are formulated as follows:
\bea
\label{active mass 2}
m_1 &=& \frac{(v_up)^2}{M_1}=
     \frac{2m_0}{\left|3-2\tilde{\kappa}\sqrt{1+\epsilon+\epsilon^2}\right|}= \frac{2m_0}{\sqrt{(3-2\kappa\sqrt{1+\epsilon+\epsilon^2}c_\phi)^2+(2\kappa\sqrt{1+\epsilon+\epsilon^2}s_\phi)^2}}\nn\\
m_2 &=& \frac{(v_up)^2}{M_2} =
\frac{m_0}{\left|(2+\epsilon)\tilde{\kappa} \right|}= \frac{m_0}{(2+\epsilon)\kappa},\\
m_3 &=& \frac{(v_up)^2}{M_3} =
\frac{ 2m_0}{\left|3+2\tilde{\kappa}\sqrt{1+\epsilon+\epsilon^2}\right|} =  \frac{2m_0}{\sqrt{(3+2\kappa\sqrt{1+\epsilon+\epsilon^2}c_\phi)^2+(2\kappa\sqrt{1+\epsilon+\epsilon^2}s_\phi)^2}},\nn
\eea
where
\begin{equation}\label{eq_M0relation}
m_0 = \frac{(v_up)^2}{M_0}
\end{equation}  is real and positive.

As we can see, since the charged lepton matrix is diagonal, the lepton mixing matrix $U_{\rm PMNS}$ is, apart from the diagonal Majorana CP-violating phase matrix $U'_P$, exactly the neutrino mixing matrix $U_\nu$, namely
\bea
U_{\rm PMNS} &\equiv&  U_\nu = U_R^\ast = U_{\rm TB}U_1^\ast U^\ast_P\\
    &=& e^{i\varphi_1} \left( \begin{array}{*{20}{c}}
\sqrt{\frac{2}{3}}c_{\theta} & \sqrt{\frac{1}{3}} & \sqrt{\frac{2}{3}}s_{\theta} e^{-i\zeta} \\
 \frac{-c_{\theta}}{\sqrt{6}}+\frac{s_{\theta}}{\sqrt{2}}e^{i\zeta} &  \sqrt{\frac{1}{3}}&  \frac{-c_{\theta}}{\sqrt{2}}-\frac{s_{\theta}}{\sqrt{6}}e^{-i\zeta}\\
 \frac{-c_{\theta}}{\sqrt{6}}-\frac{s_{\theta}}{\sqrt{2}}e^{i\zeta} &  \sqrt{\frac{1}{3}}&  \frac{c_{\theta}}{\sqrt{2}}-\frac{s_{\theta}}{\sqrt{6}} e^{-i \zeta}
\end{array} \right)
\left( \begin{array}{*{20}{c}}
 1  & 0  &   0 \\
 0  & e^{i(\varphi_2-\varphi_1)/2} & 0 \\
 0  & 0  & e^{i(\varphi_3-\varphi_1)/2}
\end{array} \right)\label{eq_UPMNS0}
\eea
Comparing with the standard parametrization of $U_{\rm PMNS}$~\cite{Tanabashi:2018oca},
\begin{eqnarray}
U_{\rm PMNS}& =& {\left(\begin{array}{ccc}
	c_{12}c_{13} &  s_{12}c_{13}& s_{13}e^{-i\delta}\\
	-c_{23}s_{12}-s_{23}c_{12} s_{13}e^{i\delta} & c_{23}c_{12}-s_{23}s_{12} s_{13}e^{i\delta} & s_{23}c_{13} \\
	s_{23}s_{12}-c_{23}c_{12} s_{13}e^{i\delta} & -s_{23}c_{12}-c_{23}s_{12} s_{13}e^{i\delta} & c_{23}c_{13}
	\end{array}\right)}U'_P,  \label{eq_UPMNSg}\\
U'_P &=& {\rm diag}(1,~e^{i\alpha_{21}/2}, ~e^{i\alpha_{31}/2}),
\end{eqnarray}
where $c_{ij}=\cos\theta_{ij},\ s_{ij}=\sin\theta_{ij}~ (ij = 12,~23,~13)$, $\delta$ and $\alpha_{21}, \alpha_{31}$ are the Dirac and two Majorana CP-violating phases, respectively. Then
we can derive, up to the second order of $s_{13}$, namely $\mathcal{O}(s^2_{13})$,  the lepton mixing angles and CP phases as
\bea
\label{mixing angles}
s_{13} &=&|U_{e3}| = \sqrt{\frac{2}{3}}s_\theta,\nn\\
\delta &=& \zeta,~~~\alpha_{21} = \varphi_2 - \varphi_1,~~~\alpha_{31} = \varphi_3 - \varphi_1,\nn \\
s^2_{12} &=& \frac{|U_{e2}|^2}{1-|U_{e3}|^2} ~=~\frac{1}{3(1-s^2_{13})},\\
s^2_{23} &=& \frac{|U_{\mu 3}|^2}{1-|U_{e3}|^2}~\simeq~ \frac{1}{2}+ \frac{1}{\sqrt{2}}\left(s_{13}-\frac{3}{4}s^2_{13}\right)c_\delta,\nn
\eea
where the mixing angle $\theta$ is obtained as
\be
 t_{2\theta} = \frac{\sqrt{3}\epsilon\tilde{\kappa}}{(2+\epsilon)\tilde{\kappa}\cos\zeta-3i \sin\zeta}.
\ee
Apart from that, the consistency of the imaginary parts between all elements of the two matrices in Eqs. \eqref{eq_UPMNS0} and \eqref{eq_UPMNSg} results in $\sin\delta=\sin\zeta=0$, up to the order $\mathcal{O}(s^2_{13})$. Following this, we hereafter take $\zeta= \pi$ (therefore the Dirac CP phase $\delta =\pi$ which is consistent with its experimental values at $3\sigma$ given in  Ref. \cite{Tanabashi:2018oca}), and then we get
\be
\label{theta}
 t_{2\theta} = \frac{-\sqrt{3}\epsilon}{2+\epsilon}.
\ee
As a result, $\epsilon$ can be written as a function of  $s_{13}$, namely
\begin{equation}\label{eq_vareps13}
	\epsilon = -\frac{4 t_{\theta }}{t_{\theta } \left(2-\sqrt{3} t_{\theta }\right)+\sqrt{3}},
\end{equation}
where $t_{\theta}=s_{\theta}/c_{\theta}$ and $s_{\theta}=\sqrt{3/2}s_{13}$.
We note that $s_{13}$ is real, hence $\epsilon$ is real too. Combining with the notation given in Eq.~\eqref{MR_phases}, the complex parameter $\tilde{\kappa}$ can be written as
\begin{equation}\label{eq_kappa}
\tilde{\kappa}=\kappa \times e^{i\phi},\quad \kappa=|\tilde{\kappa}|>0.
\end{equation}
Before coming to the numerical investigation, we would like to give some interesting comments to distinguish our work from previous works.  Although the structures of  $m_D$ and $M_R$ introduced in our model are slightly different from those  in the model given by Rref.~\cite{Karmakar:2014dva}, the two models have the same small mixing angle $\theta$ that has the same relation with $s_{13}$, namely Eq.~\eqref{mixing angles},  leading to similar formulas for $s^2_{12}$ and $s^2_{23}$ up to the order  $\mathcal{O}(s^2_{13})$. In addition, $\theta$ is determined by Eq.~\eqref{theta}, which seems to have the same form as given in Ref.~\cite{Karmakar:2014dva}, where $\epsilon=-\lambda_1$. However, the important difference is that we have   proved that $\epsilon$ is real based on  Eq.~\eqref{eq_vareps13}, while $\lambda_1$ is  the  modulus of a more general complex parameter. This means that we will scan fewer independent parameters in our numerical calculation. Another difference is that, while $\delta=0$ is chosen without explanation in Ref.~\cite{Karmakar:2014dva}, we  obtain  $\delta=\pi$ from the condition $s_{\delta}=0$  mentioned above and the recent neutrino oscillation data.

Based on Eqs. (\ref{active mass 2}), (\ref{mixing angles}), and \ref{theta}), the three active neutrino masses, the three mixing angles and the Dirac CP phase are explicitly shown in terms of the model parameters $ m_0,\epsilon, \kappa$, and $\phi$.  At present, we have five experimental results that are taken as inputs in our numerical analysis, given at $3\sigma$ by Ref. \cite{Tanabashi:2018oca} for the normal hierarchy (NH) of the active neutrino mass spectrum as
\bea
\label{LowE data}
s^2_{12} &=& 0.250-0.354;~~\Delta m_{21}^2 (10^{-5}{\rm eV}^2) = 6.93-7.96, \nonumber\\
\Delta m_{31}^2(10^{-3}{\rm eV}^2) &=& 2.45-2.69, ~~s^2_{13}  = 0.0190-0.0240, \\
s^2_{23} &=& 0.381-0.615,~~\delta/\pi (2\sigma) = 1.0-1.9,\nn
\eea
and for the inverted hierarchy (IH) as ($\theta_{12}$ and $\Delta m_{21}^2$ are unchanged)
\bea
\label{LowE data 1}
\Delta m_{23}^2(10^{-3}{\rm eV}^2) &=& 2.42-2.66, ~~s^2_{13}  = 0.0190-0.0242,\nn\\
s^2_{23} &=& 0.384-0.636,~~\delta/\pi (2\sigma) = 0.92-1.88.
\eea
where $\Delta m_{ij}^2 = m_i^2 - m_j^2$.
%%%%%%%%%%%%%%%%%%%%%%%%%%%%%%%%%%%%%%%%%%%%%%%%%%%

\begin{figure}[ht]
	\centering
	\minipage{0.485\textwidth}
	\includegraphics[width=7.5cm]{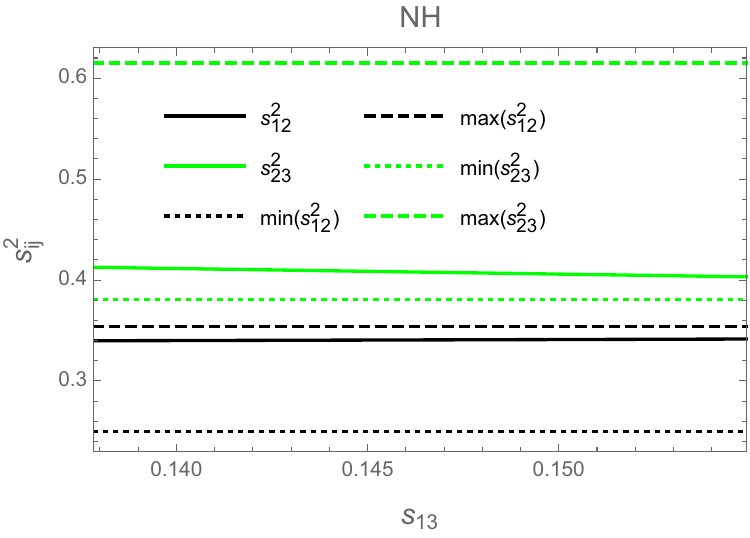}
	\endminipage
	\hfill
	\quad
	\minipage{0.485\textwidth}
	\includegraphics[width=7.5cm]{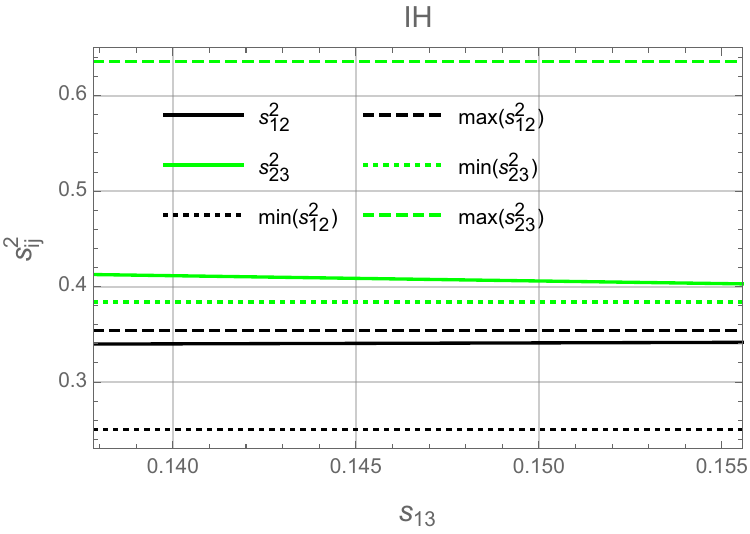}
	\endminipage
	\hfill
	\caption{ $s^2_{12}$ and  $s^2_{23}$ as functions of $s_{13}$ in the left (right) panel for the NH (IH) case.}\label{fig_s2ij}
\end{figure}

\begin{figure}[ht]
	\minipage{0.485\textwidth}
	\centering
	\includegraphics[width=10cm]{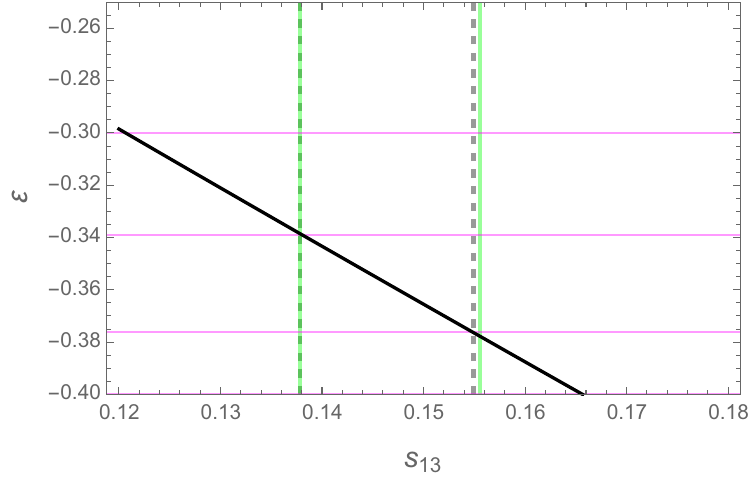}
	\endminipage
	\caption{ Plot of  $\epsilon$ as a function of $s_{13}$. The two black dashed (blue) vertical lines show the lower and upper bounds of the 3 $\sigma$ ranges of $s_{13}$  corresponding to the NH (IH) case. }\label{fig_vareps}
\end{figure}

We impose the current experimental data of active neutrino masses and mixing angles on the above relations and scan  the whole  parameter space including the following parameters: $m_0, \epsilon, \kappa, \phi$. In particular, our investigation  will try to find the allowed regions of the parameter space that satisfy both the recent experimental data of neutrino oscillation  and  leptogenesis.  Before scanning all allowed regions satisfying 3 $\sigma$ data of neutrino oscillation,  we will estimate the scanning ranges of these parameters by fixing $s_{13}$ and $\Delta m^2_{21}$ at their best-fit values, while formulating all the other  required parameters as functions of $\kappa$ and $\phi$.  Because $s^2_{12}$, $s^2_{23}$, and $\epsilon$  can be formulated as functions of only $s_{13}$, we will investigate them under constraints of recent experimental data of neutrino mixing given in Eqs.~\eqref{LowE data} and ~\eqref{LowE data 1}.  This helps us estimate the allowed ranges of these dependent parameters for further investigation. Plots of $s^2_{12}$ and $s^2_{23}$ as functions of $s_{13}$ in the 3 $\sigma$ ranges are shown in Fig.~\ref{fig_s2ij}, where the dotted and dashed lines show the respective lower and upper bounds of the 3 $\sigma$ allowed ranges given from  experimental data. With $s_{13}$ in the 3 $\sigma$ range, all values of $s^2_{12}$ and $s^2_{23}$ evaluated from the functions given in Eqs.~\eqref{LowE data} and ~\eqref{LowE data 1} always satisfy the 3 $\sigma$ allowed ranges. Hence  it is enough to pay attention only to the 3 $\sigma$ constraint of $s_{13}$. The allowed values of $\epsilon$ are presented in Fig.~\ref{fig_vareps}.

For  $s_{13}^2$ in the 3 $\sigma$ range, it is easy to derive the allowed values of $\epsilon$,
\begin{align}
\label{eq_varepsValues}
\mathrm{NH:}&\quad -0.376\leq \epsilon \leq -0.339,\crn
\mathrm{IH:}&\quad -0.378\leq \epsilon \leq-0.339.
\end{align}
At the best-fit point we have $\epsilon= -0.358\ (-0.359)$ for the NH (IH) case.

\begin{figure}[ht]
	\minipage{0.485\textwidth}
	\centering
	\includegraphics[width=8cm]{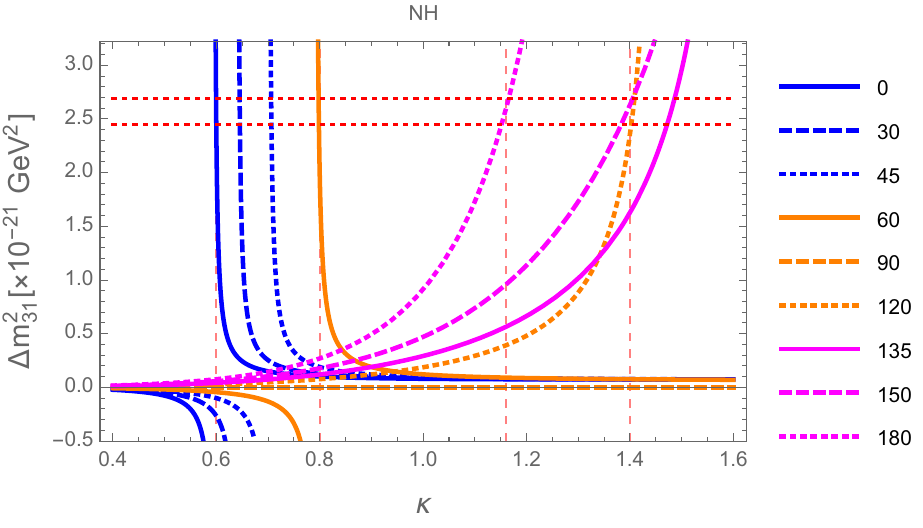}
	\endminipage
	\hfill
	\quad
	\minipage{0.485\textwidth}
	\centering
	\includegraphics[width=7cm]{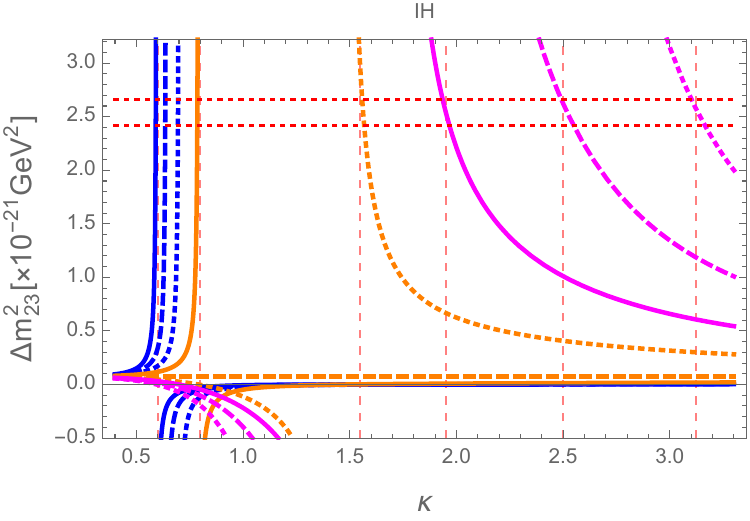}
	\endminipage
	\hfill
	\caption{ The left (right) panel presents $\Delta m^2_{31}$ ($\Delta m^2_{23}$) as a function of $\kappa$ with different fixed $0^\textrm{o}\leq\phi\leq 180^\textrm{o}$ in the NH (IH) case. The two red dotted lines show the 3 $\sigma$ allowed range of $\Delta m^2_{31}$ ($\Delta m^2_{23}$) corresponding to the NH (IH) case.}\label{fig_Dm2ijk}
\end{figure}

To estimate the allowed range of $\kappa$ and $\phi$, we use Eq.~\eqref{active mass 2} to derive $m_0$ as a function  of $\Delta m^2_{21}$, $s_{13}$ and $\tilde{\kappa}$,
\begin{align}
\label{eq_fm0}
m_0^2=\Delta m^2_{21} \left[ \frac{1}{\left|\tilde{\kappa}(2 +\epsilon) \right|^2} -\frac{1}{\left|3/2-\tilde{\kappa}\sqrt{1+\epsilon+\epsilon^2}\right|^2}\right]^{-1},
\end{align}
where $\epsilon$ and $\tilde{\kappa}$ are given by Eqs.~\eqref{eq_vareps13} and \eqref{eq_kappa}, respectively.  Inserting this form of $m_0$ into the equations in Eq. \eqref{active mass 2}, we derive $m_{1,2,3}$, $\Delta m^2_{31}$, and $\Delta m^2_{23}$ as functions of  $\Delta m^2_{21}$,  $s_{13}$, $\kappa$, and $\phi$. Using the best-fit values of  $\Delta m^2_{21}$ and $s_{13}$, we can plot $\Delta m^2_{31}$ ($\Delta m^2_{23}$) as functions of $\kappa$ with different fixed $\phi$.  In  Fig.~\ref{fig_Dm2ijk},  $\Delta m^2_{31}$ and $\Delta m^2_{23}$ corresponding to the two NH and IH cases are plotted as functions of $\kappa$ with different fixed  $\phi$.

\begin{figure}[ht]
	\minipage{0.485\textwidth}
	\centering
	\includegraphics[width=7.5cm]{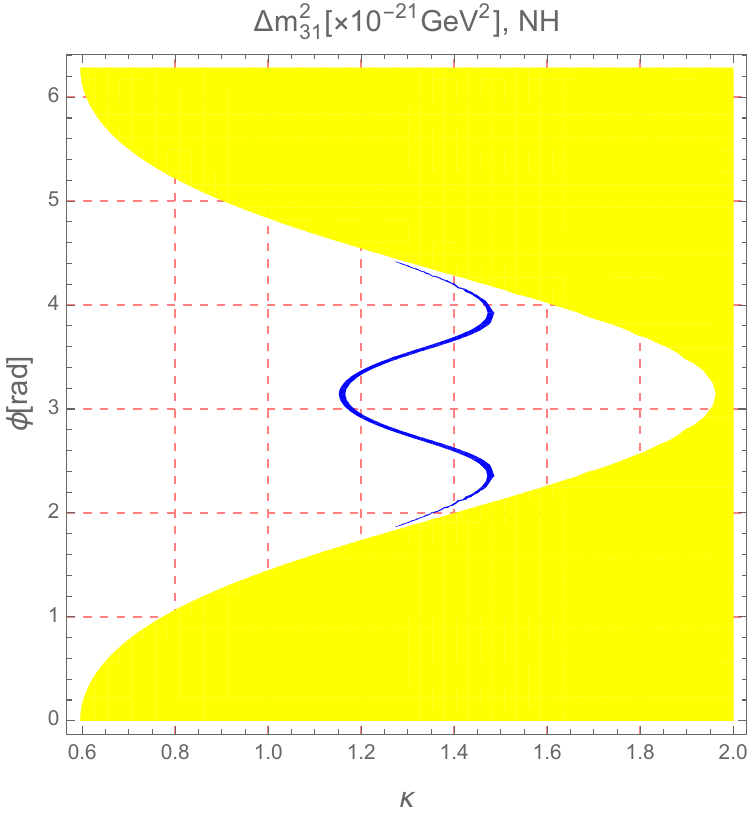}
	\endminipage
	\hfill
	\quad
	\minipage{0.485\textwidth}
	\centering
	\includegraphics[width=7.5cm]{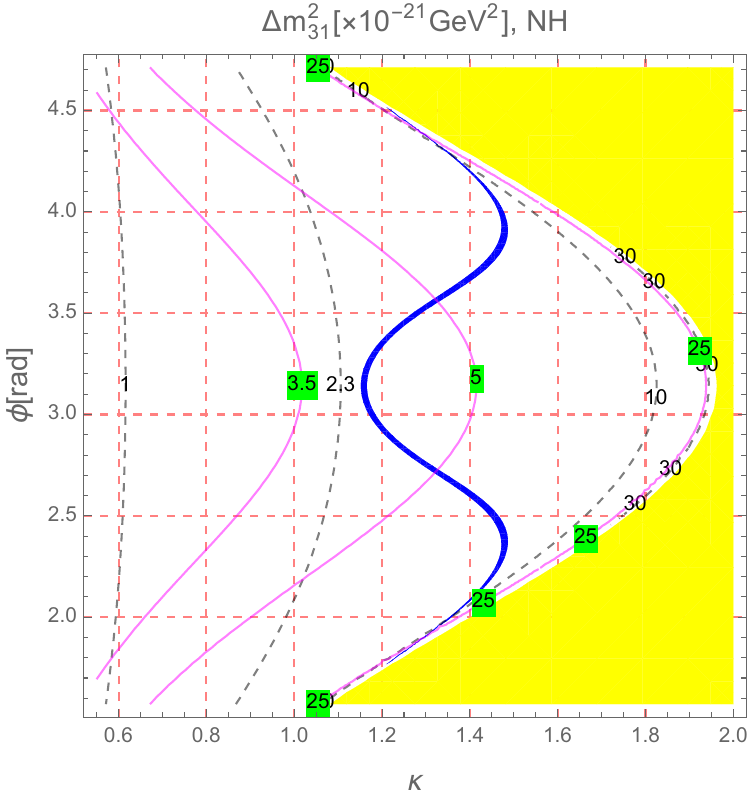}
	\endminipage
	\hfill
	\caption{ Contour plots of  $\Delta m^2_{31}$ as functions of $\kappa$ and $\phi$ in the NH case.  The blue regions  show the 3 $\sigma$ allowed range of $\Delta m^2_{31}$. The yellow regions are excluded by the condition $\Delta m^2_{21}/m_0^2>0$. The magenta and black dashed curves in the right panel show the respective constant values of $m_{0}\times 10^{11}\ [\mathrm{GeV}]$ and $m_{1}\times 10^{12}\ [\mathrm{GeV}]$. }\label{fig_contourd2mij}
\end{figure}

 Here we chose the plot range of $\phi$ as  $0^\textrm{o}\leq \phi\leq 180^\textrm{o}$ because the results in the range  $180^\textrm{o}\leq \phi\leq 360^\textrm{o}$ are repeated. We see that with every fixed $\phi$, the respective allowed range of $\kappa$ is very narrow. In addition, the two values of $\phi=90^\textrm{o}$ and $270^\textrm{o}$ are  ruled out completely for all $\kappa$  because they always result in $m_1=m_3$, leading to $\Delta m^2_{31}=0$ and $\Delta m^2_{21}=\Delta m^2_{23}$ ruled out by both the NH and the IH data.  The allowed regions divide into three, namely $\kappa<1$ for $0^\textrm{o}\leq \phi<90^\textrm{o}$ and  $270^\textrm{o}<\phi<360^\textrm{o}$, while  $\kappa>1$ for $90^\textrm{o}<\phi<270^\textrm{o}$.  In fact, the allowed regions are more strict because they must satisfy an additional condition that the formula of $m_0^2$ given in Eq.~\eqref{eq_fm0} is positive. To see how the condition $m^2_0>0$ works, we use  the contour plots in Fig.~\ref{fig_contourd2mij} for the NH case, where $\Delta m^2_{31}$, $m^2_0$, $m_0$, and $m_1$ are functions of $\phi$ and $0.6\leq \kappa<2$, which is derived from the allowed $\kappa$ shown in Fig.~\ref{fig_Dm2ijk}.

We can see in Fig.~\ref{fig_contourd2mij} that the allowed region is divided into two symmetric subregions by the horizontal axis $\phi=180^\textrm{o}$, as mentioned previously. In addition, in each subregion, for example the allowed region with $90^\textrm{o} <\phi <180^\textrm{o}$, there exists another symmetric horizontal axis $\phi=135^\textrm{o}$ where two values of $\phi=135^\textrm{o} \pm x^\textrm{o}$ will give the same $\Delta m^2_{31}$ for one fixed $\kappa$. Hence, in Fig.~\ref{fig_Dm2ijk} the two lines $\phi=120^\textrm{o}$ and $\phi=150^\textrm{o}$ result in the same  $\Delta m^2_{31}$, and the line $\phi=120<135^\textrm{o}$ is different from the three others lines $\phi=180^\textrm{o},\ 150^\textrm{o},\ 135^\textrm{o} \ge135^\textrm{o}$.

Now, only the region satisfying  $1.15<\kappa<1.5$ and $90^\textrm{o}<\phi<270^\textrm{o}$ are allowed for the NH case. We also roughly estimate the allowed ranges of $m_0$ and the lightest active neutrino mass $m_1$ as $0.035\ \mathrm{ eV}<m_0<0.25\ \mathrm{eV}$ and $0.002\ \mathrm{ eV}<m_1<0.03\ \mathrm{eV}$.

\begin{figure}[ht]
	\begin{center}
		\minipage{0.485\textwidth}
		\includegraphics[width=7.cm]{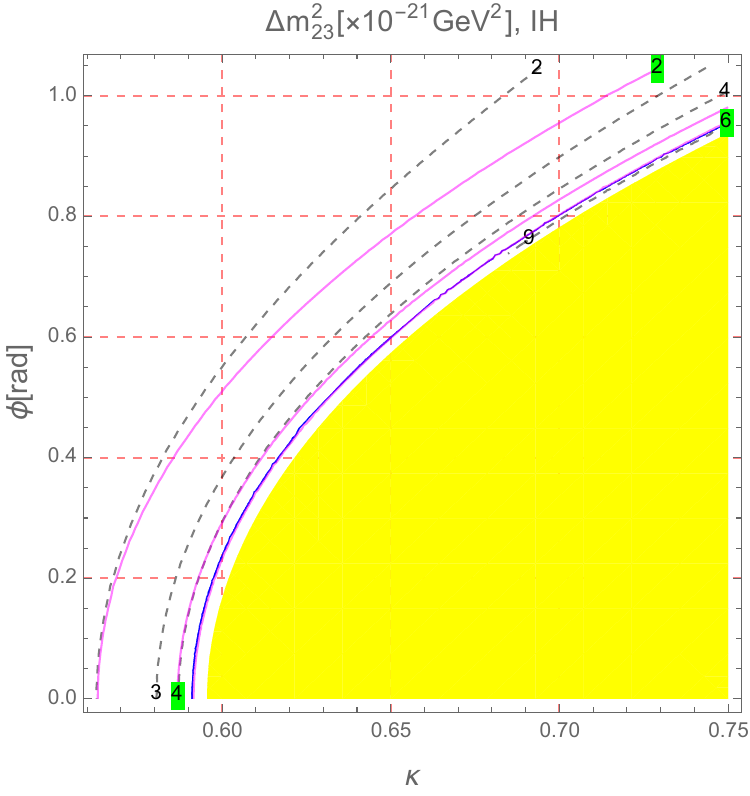}
		\endminipage
		\hfill
		\quad
		\minipage{0.485\textwidth}
		\includegraphics[width=7.cm]{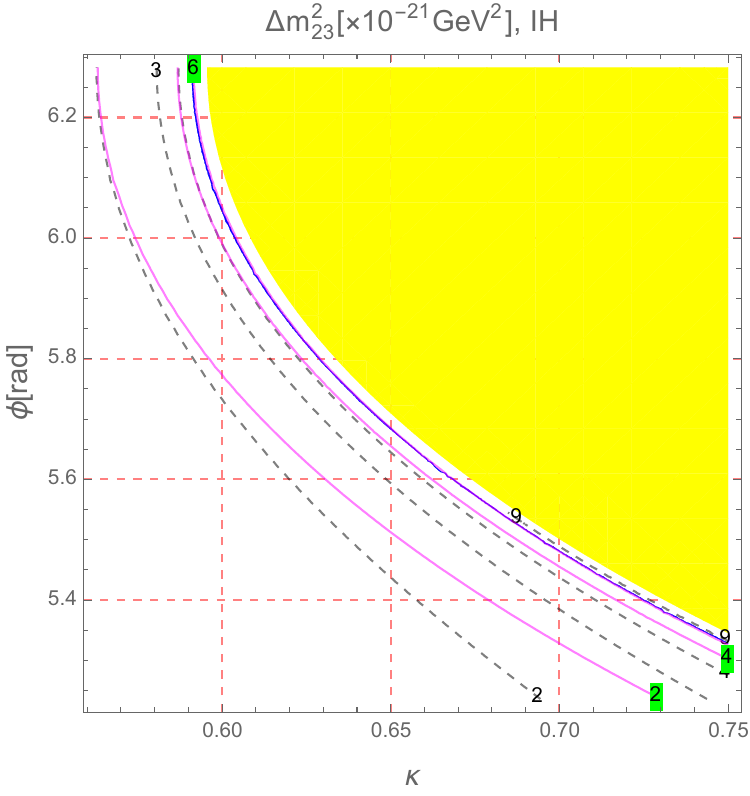}
		\endminipage
		\hfill
	\end{center}
	\caption{Contour plots of  $\Delta m^2_{23}$  as functions of $\kappa$ and $\phi$ in the range $0\leq\phi\leq\pi/3$ (left) and $5\pi/3\leq\phi\leq2\pi$ for  the IH case.  The blue regions   show the  3 $\sigma$ allowed range of $\Delta m^2_{23}$. The yellow regions are excluded by the condition $\Delta m^2_{21}/m_0^2>0$. The magenta and black dashed curves in the right panel show respective constant values of $m_{0}\times 10^{11}\ [\mathrm{GeV}]$ and $m_{1}\times 10^{11}\ [\mathrm{GeV}]$. }\label{fig_contourd2mij1}
\end{figure}

In the IH case,  illustrations are shown in Fig.~\ref{fig_contourd2mij1}, where the contour plots are limited in two ranges of $0^\textrm{o}\leq \phi\leq 60^\textrm{o}$ and $300^\textrm{o} <\phi< 360^\textrm{o}$.

The allowed regions satisfy that $0.5<\kappa<0.8$ and $\phi \in (0^\textrm{o},90^\textrm{o})\cup (270^\textrm{o}, 360^\textrm{o})$.  Values of $\phi$ close to $90^\textrm{o}$ and $270^\textrm{o}$ are excluded.   Indeed, the total allowed regions respecting the 3 $\sigma$ allowed range of $\Delta m^2_{23}$ correspond to $0^\textrm{o}\leq \phi< 90^\textrm{o}$ and $270^\textrm{o}<\phi<360^\textrm{o}$. Crude estimations of $m_0$ and $m_1$ are $0.02\ \mathrm{ eV}<m_0<0.06\ \mathrm{eV}$ and $0.04\ \mathrm{ eV}<m_1<0.09\ \mathrm{eV}$.

Particular allowed pairs of $(\kappa,\phi)$ are collected in Tables~\ref{table_KappaPhi} and \ref{table_KappaPhiI} for the NH and the IH cases, respectively. This will be very convenient for investigating the LFV decays later.
\begin{table}[ht]
\caption{Allowed values of $(\kappa,\phi)$  generating active neutrino data in $3\sigma$ range with fixed values at best-fit points of  $s_{13}^2=0.0215$  and $\Delta  m_{21}^2 = 7.37  \times10^{-23} \text{GeV}^2$ in the NH  case. The best-fit values and   allowed ranges of $\kappa$ corresponding to every fixed $\phi$ are presented in the column two.  }\label{table_KappaPhi}
	\begin{tabular}{ccccc}
		\hline
	$\phi$  & $\kappa$  &  $\Delta  m_{31}^2 $&  $  m_{0} $&  $  m_{1} $	 \\
	~[$^\textrm{o}$]&  best-fit,\;[allowed range] & $ \left[\times10^{-21}\  \text{GeV}^2\right]$&  $  \left[\times10^{-11}\ \text{GeV}\right]$&  $   \left[\times10^{-11}\ \text{GeV}\right]$	 \\
	\hline
	95 & 1.1379,\;[1.1377, 1.1381]& 2.5785 & 22.744 & 12.143 \\
	\hline
	120 & 1.4056,\;[1.4027, 1.4087] & 2.5366 & 8.6036 & 3.6293 \\
	\hline
	135 & 1.4788, [1.4720, 1.4863] & 2.5593 & 6.0594 & 2.3432 \\
	\hline
	150 & 1.3938, [1.3834, 1.4055] & 2.5599 & 3.9809 & 1.5129 \\
	\hline
	180 & 1.1582, [1.1518, 1.1654] & 2.5594 & 2.4929 & 0.99067 \\
	\hline
	210 & 1.3938, [1.3834,  1.4055] & 2.5599 & 3.9809 & 1.5129 \\
	\hline
	225 & 1.4788, [1.4720, 1.4863] & 2.5593 & 6.0594 & 2.3432 \\
	\hline
	240 & 1.4056, [1.4027, 1.4087] & 2.5603 & 8.6447 & 3.6459 \\
	\hline
	265 & 1.1379, [1.1377, 1.1381] & 2.5601 & 22.662 & 12.099 \\
	\hline
\end{tabular}
\end{table}
In the second column of every table, the values of $\kappa$ are determined at the best-fit value and 3 $\sigma$ range of $\Delta\,m^2_{31}$ ($\Delta\,m^2_{23}$) for the NH (IH) scheme.

Comparing with previous work~\cite{Karmakar:2014dva}, we can see many new interesting results  shown in the numerical estimation here. First, in our new approach, $\epsilon$ is investigated as a real function of $s_{13}$. Consequently, both $s^2_{13}$ and  $s^2_{23}$ are also written as functions of $s_{13}$, leading to a very interesting results that these two quantities  always satisfy the 3 $\sigma$ ranges.  In addition, the constraint of  $\epsilon$  is determined precisely from the 3 $\sigma$ allowed range of $s_{13}$. This approach also shows us clearly that the two allowed regions of the pairs $(\kappa,\phi)$ corresponding to the two NH and IH cases are completely distinguished.
\begin{table}[ht]
		\caption{Allowed values of $(\kappa,\phi)$  generating active neutrino data in 3 $\sigma$ range with fixed values at best-fit points of  $0.0216$ and $\Delta  m_{21}^2 = 7.37  \times10^{-23}\ \text{GeV}^2$ in the  IH case. }\label{table_KappaPhiI}
	 \begin{tabular}{ccccc}
	 	\hline
	$\phi$  & $\kappa$  &  $\Delta  m_{23}^2$&  $  m_{0} $&  $  m_{1} $	 \\
	~[$^\textrm{o}$]&  best-fit, [allowed range] & $ \left[\times10^{-21}\ \text{GeV}^2\right]$&  $  \left[\times10^{-11}\ \text{GeV}\right]$&  $   \left[\times10^{-11}\ \text{GeV}\right]$	 \\
	\hline
	10 & 0.5960, [0.5958, 0.5962] & 2.5377 & 5.6341 & 5.6958 \\
	\hline
	30 & 0.6357, [0.6354, 0.6359] & 2.6545 & 6.3001 & 5.9753 \\
	\hline
	45 & 0.6953, [0.6950, 0.6956] & 2.5293 & 7.0243 & 6.0955 \\
	\hline
	60 & 0.7859, [0.7856, 0.7862] & 2.5262 & 8.6822 & 6.6763 \\
	\hline
	85 & 1.0207, [1.0205, 1.0208] & 2.5821 & 22.271 & 13.267 \\
	\hline
	275 & 1.0207, [1.0205, 1.0208] & 2.5821 & 22.271 & 13.267 \\
	\hline
	300 & 0.7859, [0.7856, 0.7862] & 2.5262 & 8.6822 & 6.6763 \\
	\hline
	315 & 0.6953, [0.6950, 0.6956] & 2.5293 & 7.0243 & 6.0955 \\
	\hline
	350 & 0.5960, [0.5958, 0.5962] & 2.5490 & 5.6469 & 5.7088 \\
	\hline
\end{tabular}
\end{table}

From the above discussion, we have shown that by fixing $s_{13}$ and $\Delta m^2_{21}$ we can estimate the reasonable ranges of all parameters $\epsilon$, $m_0$, $\kappa$, and $\phi$.  This is also consistent with the derivation of $m_0$ from the seesaw formula $m_0=\frac{(pv_u)^2}{M_0} \simeq \sqrt{\Delta m^2_{31(23)}} \simeq 0.05$ eV for the best-fit data.   We emphasize that, although our first approach for numerical investigation seems similar to that given in Ref.~\cite{Adhikary:2008au}, our detailed discussion added more strict conditions for $m_0^2$ and $\Delta m^2_{31,23}$ to show precisely the allowed ranges of  $\kappa$ and $\phi$.  More importantly,  in the following numerical investigation  we will  scan the parameter space including  four independent parameters $(m_0,\ \epsilon,\, \kappa,\; \phi)$ around the ranges that have been estimated above to collect all allowed points which satisfy all of the 3 $\sigma$ experimental data of the NH or the IH cases. This method of investigation is more general than those mentioned in Refs. \cite{Adhikary:2008au,Karmakar:2014dva}. Coming back to our numerical investigation,  for the NH (IH) the unknown parameters  get  random values in the following ranges: $0.02\ \textrm{eV}\leq m_0\leq 0.15 \ \textrm{eV}$ ($0.05\ \textrm{eV}\leq m_0\leq 0.15 \ \textrm{eV}$), $-0.45\leq \epsilon \leq -0.25$ ($-0.4\leq \epsilon \leq-0.32$), $1\leq \kappa \leq 1.7$ ($0.45\leq \kappa \leq 1.05$),  and $0\leq \phi \leq 2\pi$.  Finally, the RHN mass scale and $t_{\beta}$ are chosen as $M_0= 10^{10}$ GeV and $t_{\beta}=3$ for the  numerical investigation of $p$, which  is the global parameter  of the Dirac neutrino Yukawa coupling matrix $Y_\nu$ roughly estimated by $p^2 \simeq \frac{M_0 \sqrt{\Delta m^2_{31(23)}}}{v_u^2}$.

\begin{figure}[ht]
	\minipage{0.485\textwidth}
	\centering
	\includegraphics[width=6.5cm]{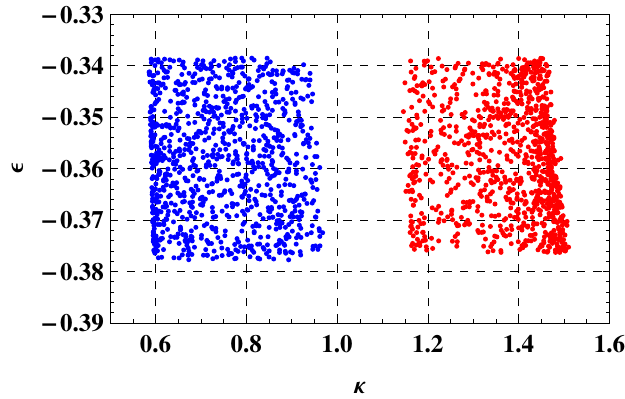}
	\caption{The allowed values of $\kappa$ and $\epsilon$ of the model. The red and blue patterns correspond to the NH and the IH of active neutrino masses, respectively.}\label{kappa-epsilon}
	\endminipage
	\hfill
	\quad
	\minipage{0.485\textwidth}
	\centering
	\includegraphics[width=6.5cm]{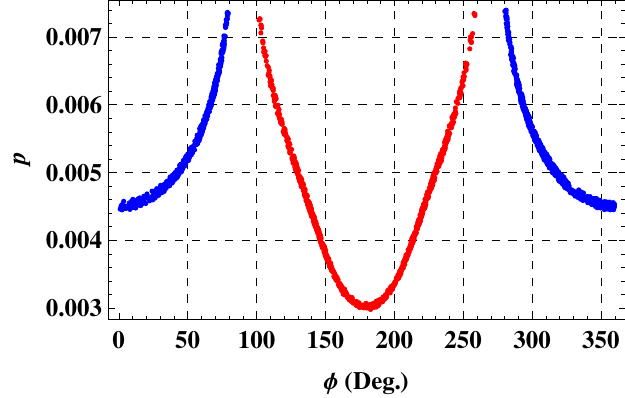}
	\caption{The correlation between the allowed values of $\phi$ and the Dirac neutrino coupling factor $p$ of the model. The roles of color patterns are the same as Fig. \ref{kappa-epsilon}.}\label{phi-p}
	\endminipage
	\hfill
\end{figure}

The parameter spaces ($\kappa, \epsilon$) and ($\phi, p$) are respectively plotted Figs \ref{kappa-epsilon} and \ref{phi-p}, where the red and blue patterns  represent the  allowed regions of NH and the IH cases, respectively. Hereafter, we continue using these conventions unless  otherwise stated. 
%%%%%%%%%%%%%%%%%%%%%%%%%%%%%%%%%%%%%%%%%%%%%%%%%
Note that $M_0$ and $t_\beta$ are absorbed into $m_0$ by the seesaw formula. As a result, the allowed regions of  the parameter spaces plotted in Fig.~\ref{kappa-epsilon} is independent from the values of $t_\beta$ and $M_0$.
%%%%%%%%%%%%%%%%%%%%%%%%%%%%%%%%%%%%%%%%%%%%%%%%%%%
\begin{figure}[ht]
	\minipage{0.485\textwidth}
	\centering
	\includegraphics[width=6.5cm]{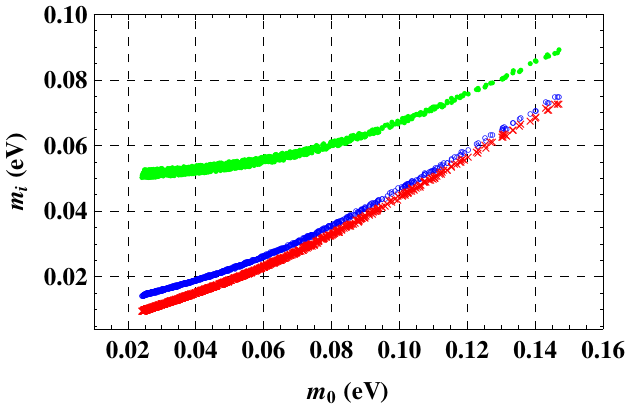}
	\caption{The active neutrino masses $m_i$ as a function of  $m_0$ for the NH case.}\label{N_mass0-massi}
	\endminipage
	\hfill
	\quad
	\minipage{0.485\textwidth}
	\centering
	\includegraphics[width=6.5cm]{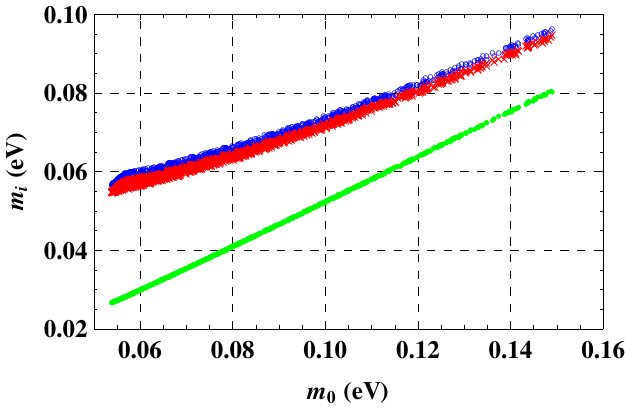}
	\caption{The active neutrino masses $m_i$ as a function of $m_0$ for the IH case.}\label{I_mass0-massi}
	\endminipage
	\hfill
\end{figure}
%%%%%%%%%%%%%%%%%%%%%%%%%%%%%%%%%%%%%%%%%%%%%%%%%
%%%%%%%%%%%%%%%%%%%%%%%%%%%%%%%%%%%%%%%%%%%%%%%%%%%

The light neutrino masses predicted by the model are respectively plotted in Figs. \ref{N_mass0-massi} and \ref{I_mass0-massi}, as functions of the light neutrino mass scale $m_0$ for the NH and IH cases. There, the red, blue, and green plots represent for $m_1,\ m_2$, and $ m_3$, respectively.
\begin{figure}[ht]
	\minipage{0.485\textwidth}
	\centering
	\includegraphics[width=6.5cm]{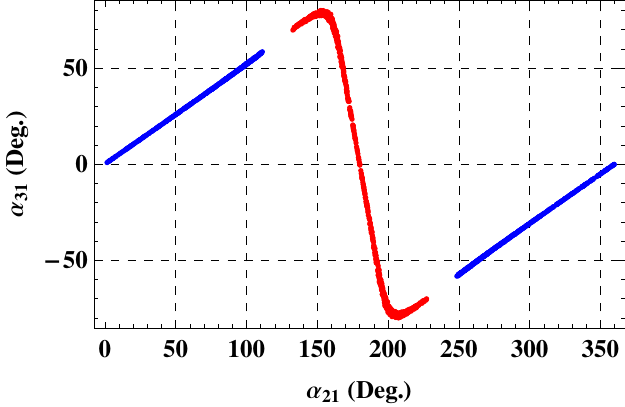}
	\caption{The predictions of the model for the Majorana CP-violating phases for the NH (red) and the IH (blue) cases.}\label{Alpha21-Alpha31}
	\endminipage
	\hfill
	\quad
	\minipage{0.485\textwidth}
	\centering
	\includegraphics[width=6.5cm]{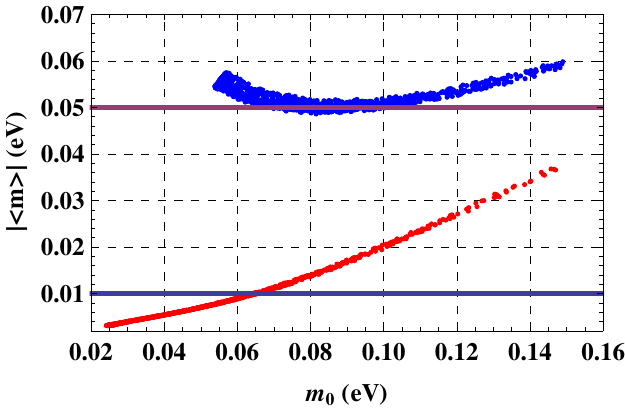}
	\caption{The predictions of the effective neutrino mass $|\langle m\rangle|$ as a function of the active neutrino mass scale $m_0$.}\label{mass0-NDBD}
	\endminipage
	\hfill
\end{figure}
%%%%%%%%%%%%%%%%%%%%%%%%%%%%%%%%%%%%%%%%%%%%%%%%%

  We can recognize that the neutrino masses are strong hierarchy with small values of $m_0$ and they can be quasi-degenerate, $m_1 \cong m_2 \cong m_3 \geq 0.1$ eV \cite{Tanabashi:2018oca}, if $m_0$ approaches above 0.15 eV. The prediction of the two Majorana CP phases is shown in Fig. \ref{Alpha21-Alpha31}.

It is worth to studying the effective neutrino mass in neutrinoless double beta decay ($0\nu\beta\beta$),  $|\langle m\rangle|$,  with the form given in Ref. \cite{Tanabashi:2018oca} as
\bea
|\langle m\rangle| &=& \left |m_1(U_{\rm PMNS})_{e1}^2+ m_2(U_{\rm PMNS})_{e2}^2 + m_3(U_{\rm PMNS})_{e3}^2\right|\nonumber\\
&=& \left|\Big(m_1 c^2_{12}+ m_2 s^2_{12}e^{i\alpha_{21}}\Big) c^2_{13} + m_3 s^2_{13}e^{i(\alpha_{31}-2\delta)}\right|.
\eea
The prediction of the effective mass $|\langle m\rangle|$ is plotted in Fig. \ref{mass0-NDBD} as a function of the lightest active neutrino mass $m_0$ for the NH (red plot) and IH (blue plot) cases. In this figure, the two horizontal lines are the prospect bounds for $|\langle m\rangle|$ of a new generation of $0\nu\beta\beta$ experiments \cite{Tanabashi:2018oca}. Numerically, our predictions of $|\langle m\rangle|$ turn out to be 0.002 eV $\leq |\langle m\rangle| \leq$ 0.038 eV for NH and 0.048 eV $\leq |\langle m\rangle| \leq$ 0.058 eV for IH.
Notice that the results from $0\nu2\beta$ by KamLAND-Zen \cite{Asakura:2014lma} and EXO-200 \cite{Albert:2014awa} indicate an upper limit on the effective neutrino mass parameter $|\langle m\rangle| $ that  $ |\langle m\rangle| \leq  (0.14-0.28)$ eV at $90\%$ CL. and  $ |\langle m\rangle| \leq (0.19-0.45)$ eV at $90\%$ CL., respectively. The most stringent upper limit now is  $|\langle m\rangle| \leq (0.061-0.165)$ eV at $90\%$ CL~\cite{KamLAND-Zen:2016pfg}.
Therefore, our result for $|\langle m\rangle|$ is still not excluded by the current experimental bounds, and we expect that our predictions for $|\langle m\rangle|$ could be measured by KamLAND-Zen and other $0\nu 2\beta$ decay experiments in their new phase which have been taking data since mid 2017; see, for the present status and future prospects, Ref.  \cite{Maneschg:2017mzu}. The future sensitivity can reach $|\langle m\rangle| =0.01$ eV; see  a summary in Ref.~\cite{Abada:2018qok}, where the sensitivity of many ongoing and planned  $0\nu\beta\beta$ experiments \cite{Auger:2012ar,Albert:2014awa,Azzolini:2018dyb,Tosi:2014mya, Obara:2017ndb, Agostini:2017iyd, Phillips:2011db, Abgrall:2017syy, Aguirre:2014lua, Artusa:2014lgv, Hartnell:2012qd, Barabash:2011aa, Karki:2018rhc, Gomez-Cadenas:2013lta} were listed. Because the two ranges of  $|\langle m\rangle|$ predicted by the NH and IH cases are completely distinguished,  $|\langle m\rangle|$  is  an  important channel to confirm experimentally the NH or the IH property once the effective mass $|\langle m\rangle|$ is measured. In addition, we can pin down the light neutrino mass scale $m_0$ and either of the active neutrino masses.

To finish the numerical investigation, we conclude some important constraints on the model parameters.  The allowed ranges of the four parameters $(m_0,\epsilon, \kappa,\phi)$ are constrained as follows.  The allowed regions for the NH case are: $0.02\,\mathrm{eV}<m_0<0.15 \,\mathrm{eV}$, $-0.038<\epsilon<-0.0345$, $1.15<\kappa<1.5$, and $90^\text{o}<\phi<270^\text{o}$.  The allowed regions for the IH case are: $0.05\,\mathrm{eV}<m_0<0.15\,\mathrm{eV}$, $-0.038<\epsilon<-0.0345$, $0.55<\kappa<1$,  $0^\text{o}<\phi<90^\text{o}$, and $270^\text{o}<\phi<360^\text{o}$.  In addition, the allowed region of the light neutrino mass scale $m_0$ lead to upper bounds of $M_0$ obtained from the perturbative limit: $M_0=(v_u^2 p)/m_0\leq 174^2\times 4\pi/(0.02\times 10^{-11})\sim O(10^{16})$ GeV.  This is consistent with the GUT scale mentioned in this work.

 Interestingly enough, in the next section we would like to study  how the BAU can  be explained by the leptogenesis scenario of the current model under the allowed regions of the parameter space discussed in this section.
%%%%%%%%%%%%%%%%%%%%%%%%%%%%%%%%%%%%%%%%%%%%%%%%%%%%%%%%%%%%%%%%%%%%%%%%%%%%%%%%%%%%
\section{\label{sec_ Leptogenesis} Leptogenesis}
We now consider how leptogenesis can work in our scenario.  The relations between heavy RHNs and active neutrino masses  are derived directly from Eq.~\eqref{eq_mnu1},
\bea
\label {Diagonal MR}
U_{R}^{\textrm{T}} M_R U_{R} = {\rm diag}(M_1,\ M_2,\ M_3) = (v_up)^2~{\rm diag}\left(\frac{1}{m_1},\frac{1}{m_2},\frac{1}{m_3}\right),\label{fVR}
\eea
where $U_{R} = U^\ast_{\rm PMNS}$ was determined precisely in the previous section. In the mass basis of the RHNs, the Dirac neutrino Yukawa coupling matrix is modified to be
\bea
Y_\nu' = U_R^{\textrm{T}}Y_\nu =U_{\rm PMNS}^\dag Y_\nu \Rightarrow ~ H=Y_\nu' Y_\nu'^\dag ~=~ p^2\times \textbf{1}.
\label{Ynup}
\eea

We study the case of flavored leptogenesis, the CP asymmetry in the decay of RHN $N_i$ to lepton flavor $l_\alpha ~ (\alpha = e, \mu, \tau)$ is defined as \cite{Abada:2006ea,Blanchet:2006be,Antusch:2006cw, Pascoli:2006ie,Pascoli:2006ci,Branco:2006ce,Branco:2006hz}
\bea
\varepsilon_{i}^\alpha &=& \frac{\Ga(N_{i}\rightarrow l_\alpha\varphi)
	-\Ga(N_{i}\rightarrow \overline{l}_\alpha\varphi^{\dag})}{\sum_{\alpha}[\Ga(N_{i}\rightarrow l_\alpha\varphi)
	+\Ga(N_{i}\rightarrow\overline{l}_\alpha\varphi^{\dag})]}\nn\\
   &=&
\frac{1}{8\pi  H_{ii}}\sum_{j\neq i}\Big\{{\rm Im}\Big[H_{ij}(Y'_\nu)_{i\alpha}(Y'_\nu)^\ast_{j\alpha} \Big]f\Big(\frac{M^{2}_{j}}{M^{2}_{i}}\Big),
\label{cpasym}
\eea
where $H = Y_\nu'Y_\nu'^\dag$, and
$M_i$ denotes the RHN masses. The loop function $f(x)$ containing the vertex and self-energy corrections is given as
\begin{equation}
f(x)=\sqrt{x}\Big[(1+x){\rm ln}\frac{x}{1+x}+\frac{2-x}{1-x}  \Big].
\label{Loop correction}
\end{equation}
%%%%%%%%%%%%%%%%%%%%%%%%%%%%%%%%%%%%%%%%%%%%%%%%%%%%%%%%%%%%%%%%%%%%%%%%%%%

Notice from Eq.(\ref{cpasym}) that, in the original model, the CP asymmetry is zero due to the fact that the Hermitian matrix $H$ is proportional to the unit matrix, see Eq. (\ref{Ynup}), and a non-vanishing CP asymmetry requires ${\rm Im}[H_{ij}({Y'}_\nu)_{i\alpha}({Y'}_\nu)_{j\alpha}^\ast]\neq 0$. Therefore,  to have leptogenesis we need to induce a non-vanishing ${H}_{ij}\ (i\neq j)$ at the leptogenesis scale. Indeed, this happens in the model under consideration because of the RG (renormalization group) effects, discussed in detail below. The RG equation for the Dirac neutrino Yukawa coupling can be written as \cite{Casas:1999tp,Chankowski:2001mx, Antusch:2002rr,Branco:2005ye,Nguyen:2012zza}
 \begin{eqnarray}
     \frac{d {Y}_{\nu}}{dt} &=&
   {Y}_{\nu}\left[\left(T-\frac{3}{4}g^{2}_2-\frac{9}{4}g^{2}_{1}\left)
   -\frac{3}{2}\right({Y}^{\dag}_{l}{Y}_{l}-{Y}^{\dag}_{\nu}{Y}_{\nu}\right)\right],
  \label{RG 2}
 \end{eqnarray}
where
$T=\mathrm{Tr}(3Y^{\dag}_{u}Y_{u}+3Y^{\dag}_{d}Y_{d}+{Y}^{\dag}_{\nu}{Y}_{\nu}+{Y}^{\dag}_l{Y}_l)$,
$Y_{u, d}$ and ${Y}_{l}$ are the Yukawa couplings of up-type and down-type
quarks and charged leptons, $g_{2,1}$ are the ${\rm SU(2)}_{L}$
and ${\rm U(1)}_{Y}$ gauge coupling constants, respectively, $ t  = \frac{1}{16\pi^{2}}\ln(M/\Lambda')$, and $M$ is an
arbitrary renormalization scale. The cutoff scale $\Lambda'$ can
be regarded as the $G_f$ breaking scale $\Lambda'=\Lambda$ and is
assumed to be of the order of the GUT scale, $\Lambda' \sim 10^{16}$
GeV.

As the structure of ${M}_R$ changes with the evolution of
the energy scale, $U_R$ depends on the scale $\Lambda'$ too. The RG evolution of
$U_R(t)$ can be written as
\begin{equation}
\frac{d}{dt}U_R=U_RA,
\end{equation}
where $A$ is an anti-Hermitian matrix $A^\dagger =-A$ due to the
unitarity of $U_R$. The components of the $A$ matrix are given by \cite{Ahn:2006rn}
\begin{eqnarray}
  A_{ij}&=& \frac{M_j+M_i}{M_j-M_i}{\rm Re}[(Y_\nu Y_\nu^\dagger)_{ij}]
  +i\frac{M_j-M_i}{M_j+M_i}{\rm Im}[(Y_\nu Y_\nu^\dagger)_{ij}].
  \label{RG A}
 \end{eqnarray}
 The running of the RHN mass scale affects very weakly  our result so we drop it here.
The RG equation for $Y'_{\nu}$ in the basis of diagonal
${M}_{R}$ is then obtained as
\begin{eqnarray}
  \label{RG 7}
   \frac{dY_{\nu}'}{dt} &=&
  Y_{\nu}'\left[\left(T-\frac{3}{4}g^{2}_2-\frac{9}{4}g^{2}_{1}\right)-\frac{3}{2}\left({Y}^{\dag}_{l}{Y}_{l}-{Y}'^{\dag}_{\nu}{Y}_{\nu}'\right)\right]
       +A^{\textrm{T}}Y_{\nu}'.
\end{eqnarray}
Finally, we obtain the RG equation for the Hermitian matrix $H=Y'_\nu Y_\nu'^\dag$ responsible for the  leptogenesis as
 \begin{eqnarray}
    \frac{dH}{dt}&=&
  2\left(T-\frac{3}{2}g^{2}_2-\frac{9}{4}g^{2}_{1}\right)H-3Y_\nu({Y}_l^\dagger {Y}_l)Y_\nu'^\dagger +3 H^2+A^{\textrm{T}} H+HA^\ast.
  \label{RG 8}
 \end{eqnarray}
With the Hermitian matrix $H$ given in Eq. (\ref{Ynup}), up to non-zero leading contributions in the right-hand side of Eq. (\ref{RG 8}), the RG is generated from the off-diagonal terms of the $H$ matrix as
\begin{eqnarray}
  \label{radiatively induced}
   {H}_{ij}(t) &\simeq &-3 y_\tau^2({Y}_\nu')_{i3}({Y}_\nu')_{j3}^\ast \times t.
 \end{eqnarray}
The flavored CP asymmetries $\varepsilon_i^\alpha$ can then be
obtained. Notice that, in this model, the tau Yukawa coupling constant ($y_\tau$) relates to that in the SM ($y_{\tau, \textrm{SM}}$) as $y_\tau^2 = y^2_{\tau,\textrm{SM}} (1+t^2_\beta)$. This enhances the CP asymmetries as $\varepsilon_i^\alpha \sim (1+t^2_\beta)$,  we will later discuss the effect of different values of $t_\beta$ on the numerical generation of the BAU.
%%%%%%%%%%%%%%%%%%%%%%%%%%%%%%%%%%%%%%%%%%%%%%%%%%%%%%%%%%%%%%%%%%%%%%%%%%%

After the CP asymmetry in the decay of $N_i$, $\varepsilon^{\alpha}_{i}$, are calculated,
the final value of $\eta_{B}$ can be calculated by solving the flavor-dependent Boltzmann equations (BE). These describe the out-of-equilibrium processes such as the decay,
inverse decay, and scattering involving the RHNs, as well as the non-perturbative sphaleron interaction. Besides the CP asymmetries $\varepsilon^{\alpha}_{i}$, the final value of BAU also depends on the wash-out factors $K^{\alpha}_{i}$ which measure the effects of the inverse decay of Majorana neutrino $N_{i}$ into the lepton flavor $\alpha$ and scalars. The parameter $K^{\alpha}_{i}$ is defined as~\cite{Abada:2006ea}
\begin{eqnarray}
K^{\alpha}_{i}=\frac{\Gamma^{\alpha}_{i}}{H(M_{i})}=(Y'^{\dag}_{\nu})_{\alpha
	i}(Y_{\nu}')_{i\alpha}\frac{\upsilon^{2}_{u}}{m_{\ast}M_{i}},
\label{washout01}
\end{eqnarray}
where $\Gamma^{\alpha}_{i}$ is the partial decay width of $N_{i}$
into the lepton flavors and Higgs scalars; $H(M_{i})$ is the Hubble parameter at temperature $T=M_{i}$ defined as $H(M_{i})\simeq(4\pi^{3}g_{\ast}/45)^{\frac{1}{2}}M^{2}_{i}/M_{\textrm{Pl}}$, where
 $M_{\textrm{Pl}}=1.22\times10^{19}$ GeV is the Planck mass, $g_{\ast}\simeq 116$ is the
effective number of degrees of freedom of the SM with two Higgs doublets, and the equilibrium
neutrino mass $m_{\ast}\simeq10^{-3}$ eV.

Due to the flavor effects, each CP asymmetry $\varepsilon_{i}^\alpha$ contributes differently to the final formula for the baryon asymmetry as~\cite{Abada:2006ea, Ahn:2010cc,Ahn:2010nw}
\begin{eqnarray}
\label{EthaB1}
\eta_B \simeq -2 \times 10^{-2}\sum_{N_{i}}\Big[\varepsilon^{e}_{i}\kappa_i^e\Big(\frac{151}{179}K^{e}_{i}\Big)+\varepsilon^{\mu}_{i}\kappa_i^\mu\Big(\frac{344}{537}K^{\mu}_{i}\Big)+\varepsilon^{\tau}_{i}\kappa_i^\tau\Big(\frac{344}{537}K^{\tau}_{i}\Big)\Big]
\end{eqnarray}
if the RHN mass is about $M_i
\leq (1+t^2_\beta)\times 10^{9}$  GeV  where the $\mu$ and $\tau$
Yukawa couplings are in equilibrium and all the flavors are to be
treated separately. If $(1+t^2_\beta)\times 10^{9}$ GeV $\leq M_i \leq
(1+t^2_\beta)\times 10^{12}$ GeV where only the $\tau$ Yukawa
coupling is in equilibrium and  treated separately while the $e$
and $\mu$ flavors are indistinguishable, then the baryon asymmetry is obtained as
\begin{eqnarray}
\label{EthaB2}
\eta_B \simeq-2\times 10^{-2}\sum_{N_{i}}\Big[\varepsilon^{2}_{i}\kappa_i^2\Big(\frac{417}{589}K^{2}_{i}\Big)
+\varepsilon^{\tau}_{i}\kappa_i^\tau\Big(\frac{390}{589}K^{\tau}_{i}\Big)\Big],
\end{eqnarray}
where
$\varepsilon^{2}_{i}=\varepsilon^{e}_{i}+\varepsilon^{\mu}_{i}$ and
$K_i^2=K_i^e+K_i^\mu$. In Eqs. (\ref{EthaB1}) and (\ref{EthaB2}), the wash-out factors $\kappa_i^\alpha$ are defined as
\begin{eqnarray}
\label{washout}
\kappa^{\alpha}_{i}\simeq\Big(\frac{8.25}{K^{\alpha}_{i}}+\Big(\frac{K^{\alpha}_{i}}{0.2}\Big)^{1.16}\Big)^{-1}.
\end{eqnarray}
%%%%%%%%%%%%%%%%%%%%%%%%%%%%%%%%%%%%%%%%%%%%%%%%%%%

\begin{figure}[ht]
	\minipage{0.485\textwidth}
	\centering
	\includegraphics[width=6.5cm]{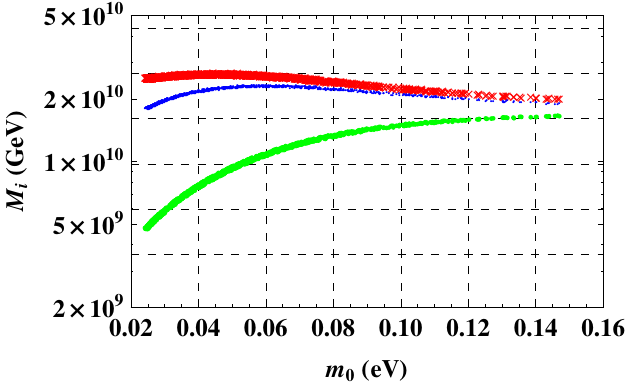}
	\caption{The RHN masses as functions of the light neutrino mass scale $m_0$ for the NH with $M_0 =10^{10}$ GeV.}\label{N_mass0-HM}
	\endminipage
	\hfill
	\quad
	\minipage{0.485\textwidth}
	\centering
	\includegraphics[width=6.5cm]{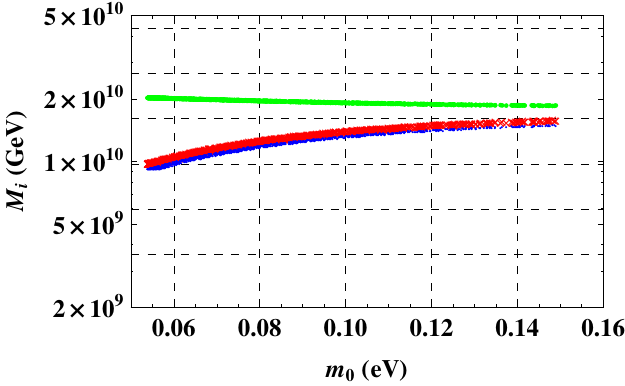}
	\caption{The RHN masses as functions of the light neutrino mass scale $m_0$ for the IH with $M_0 =10^{10}$ GeV.}\label{I_mass0-HM}
	\endminipage
	\hfill
\end{figure}
%%%%%%%%%%%%%%%%%%%%%%%%%%%%%%%%%%%%%%%%%%%%%%%%%
%%%%%%%%%%%%%%%%%%%%%%%%%%%%%%%%%%%%%%%%%%%%%%%%%%%

The allowed regions of parameter space given in Sect. II and all the formulas discussed above for $\eta_B$ are enough to allow us to investigate numerically the BAU predicted by the model under consideration. For $\tan\beta$ given in Eq.~\eqref{tbeta},   the Lagrangian in Eq. \eqref{eq_LYq} gives  $v_u> m_t/\sqrt{4\pi}$. Combining with the relation in Eq.~\eqref{SMmatching}, it can be shown easily that $\tan\beta \ge0.3$, which will be used in the following numerical investigations.  First, the mass spectra of RHN masses as functions of the active neutrino mass scale, $m_0$, are plotted in Figs. \ref{N_mass0-HM} and \ref{I_mass0-HM} for the respective NH and the IH cases, where  the red, blue, and green lines represent for $M_1,\ M_2$ and $M_3$, respectively. Those RHN masses are a strong hierarchy with small values of $m_0$ and gradually become quasi-degenerate when $m_0$ approaches values around $0.15$ eV. This enhances the generated $\eta_B$ by the so-called resonant leptogenesis \cite{Pilaftsis:2005rv}.
 \begin{figure}[ht]
 	\minipage{0.485\textwidth}
 	\centering
 \includegraphics[width=6.5cm]{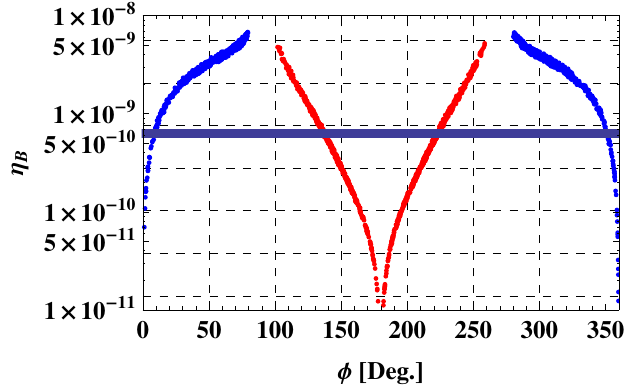}
 	\caption{The prediction of $\eta_B$ as a function of $\phi$, where $t_\beta = 3$ and $M_0 = 10^{10}$ GeV are used. The red (blue) curve represents the NH (IH) case.}\label{Phi-EtaB}
 	\endminipage
 	\hfill
 	\quad
 	\minipage{0.485\textwidth}
 	\centering
 	\includegraphics[width=6.5cm]{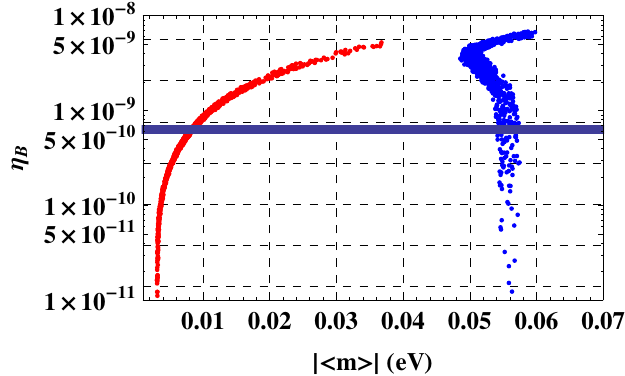}
 	\caption{The correlation between $\eta_B$ and $|\langle m\rangle|$, where  $t_\beta = 3$ and $M_0 = 10^{10}$ GeV are used. The red (blue) curve represents the NH (IH) case.}\label{NDBD-EtaB}
 	\endminipage	
 	\hfill
 \end{figure}

\begin{figure}[h]
	\minipage{0.485\textwidth}
	\centering
	\includegraphics[width=6.5cm]{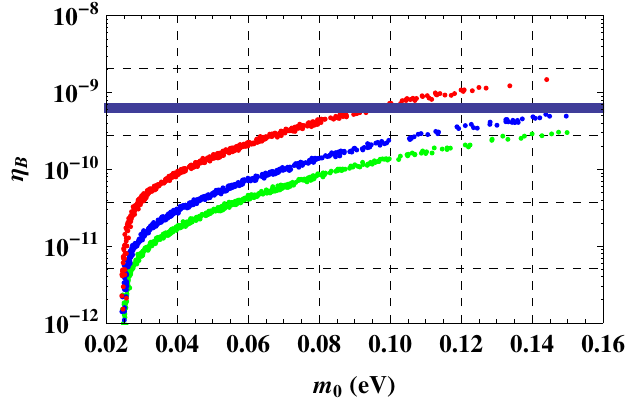}
	\caption{The prediction of $\eta_B$ for the case of NH as a function of the active neutrino mass scale $m_0$ with $M_0 = 10^8$ GeV. The green, blue and red plots correspond to $\tan\beta = 1, 3, 10$, respectively.}\label{EtaB-tanbeta-N}
	\endminipage
	\hfill
	\quad
	\minipage{0.485\textwidth}
	\centering
	\includegraphics[width=6.5cm]{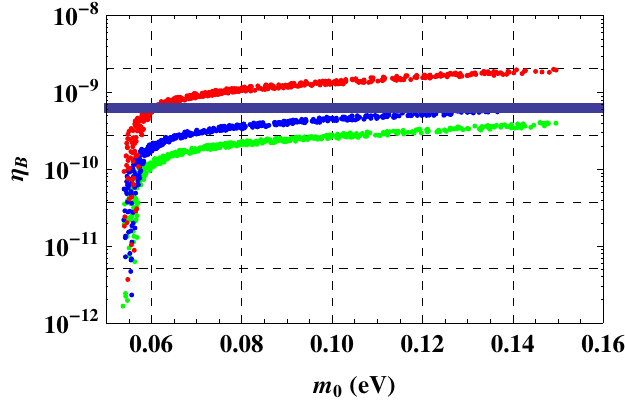}
	\caption{The prediction of $\eta_B$ for the case of IH as a function of the active neutrino mass scale $m_0$ with $M_0 = 10^8$ GeV. The green, blue and red plots correspond to $\tan\beta = 1, 3, 10$, respectively.}\label{EtaB-tanbeta-I}
	\endminipage
	\hfill
\end{figure}

 %%%%%%%%%%%%%%%%%%%%%%%%%%%%%%%%%%%%%%%%%%%%%%%%%
 Later, we can find in Figs.  \ref{EtaB-tanbeta-N} and \ref{EtaB-tanbeta-I} that $\eta_B$ increases with  increasing $m_0$ due to the effects of resonant leptogenesis.
 This is numerically proved in Fig. \ref{Phi-EtaB}, where the prediction of $\eta_B$ as a function of the phase $\phi$  is shown.
 In this figure, $\eta_B$ gets two maxima around $\phi = 90^{\textrm{o}}$ and  $\phi = 270^{\textrm{o}}$ for both cases of hierarchy of neutrino masses. The reason is that the parameter $p$ of the $Y_\nu$ matrix is proportional to $\sqrt{m_0}$, which has two maxima around $\phi = 90^{\textrm{o}}$ and  $\phi = 270^{\textrm{o}}$ for both hierarchies (see, Fig. \ref{phi-p}). Therefore, $m_0$, and hence $\eta_B$, also get their maxima around these values of the phase $\phi$. In this figure (and in Figs. \ref{NDBD-EtaB}$-$\ref{EtaB-tanbeta-I}), the solid horizontal bar represents the allowed range from  experiment for  BAU, namely  $\eta_B = (6.3 \pm 0.3)\times 10^{-10}$ \cite{Aghanim:2016yuo}.

 The correlation between $\eta_B$ and $|\langle m \rangle|$ is shown in Fig. \ref{NDBD-EtaB} for $M_0 =10^{10}$ Gev and $t_\beta = 3$, where  the red (blue) curve represents the NH (IH) case. As indicated in the previous section, once the exact value of $|\langle m \rangle|$ is confirmed we can point out the active neutrino mass scale $m_0$ and then we can find out the required values of the RHN mass in order to generate the right amount of $\eta_B$ for some given values of $t_\beta$.

%%%%%%%%%%%%%%%%%%%%%%%%%%%%%%%%%%%%%%%%%%%%%%%%%%%
%%%%%%%%%%%%%%%%%%%%%%%%%%%%%%%%%%%%%%%%%%%%%%%%%

The effects of different values of $t_\beta$ on the resultant of the $\eta_B$ are shown in Fig. \ref{EtaB-tanbeta-N} for NH and Fig. \ref{EtaB-tanbeta-I} for IH.  In these figures, the green, blue, and red plots correspond to $t_\beta = 1,\ 3$ and 10, respectively, with the mass scale of RHN $M_0 = 10^8$ GeV. We can find that $\eta_B$ increases with increasing $t_\beta$; with $t_\beta =3$ the minimum value of the mass scale $M_0$ is about $10^8$  GeV for successful leptogenesis, where the minimum values of $M_0$ for attaining the right value of $\eta_B$ are much reduced with larger values of $t_\beta$.

\begin{figure}[ht]
	\minipage{0.485\textwidth}
	\centering
	\includegraphics[width=6.5cm]{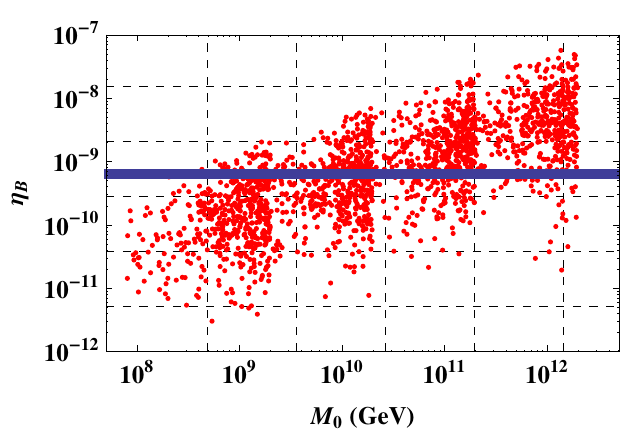}
	\endminipage
	\hfill
	\quad
	\minipage{0.485\textwidth}
	\centering
	\includegraphics[width=6.5cm]{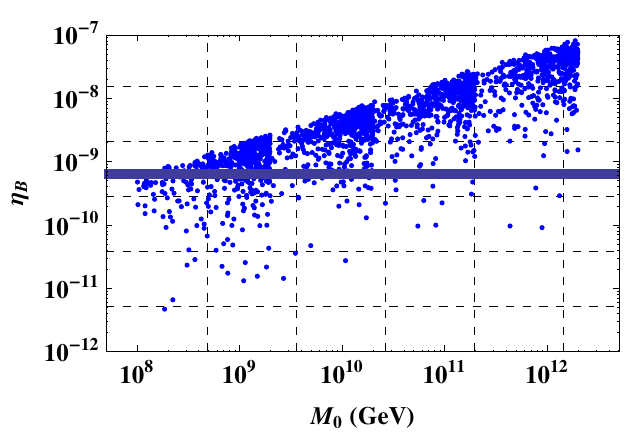}
	\endminipage
	\hfill
	\caption{The prediction of $\eta_B$ for the case of NH (IN) as a function of the RHN mass scale $M_0$  with $t_{\beta}=3$ in the left (right) panel. The allowed ranges of  $\kappa$, $\epsilon$, $\phi$, and $m_0$ summarized at	the end of Sect. \ref{sec_a4model} are  used.}\label{fig_M0-EtaB}
\end{figure}

We emphasize one important point about the constraint of the heavy neutrino mass scale.  For $1\leq t_{\beta}\leq 10$, our numerical investigation shows that the constraint is $O(10^8) \ \mathrm{GeV}\leq M_0\leq O(10^{12}) \  \mathrm{GeV}$ in order to successfully generate leptogenesis; see the illustration in Fig.~\ref{fig_M0-EtaB}.
This result can be explained by the fact that  $M_0$  relates to $m_0$, $p$, and $v_u=174s_{\beta}$ through the relation in Eq.~\eqref{eq_M0relation}, where $p\leq \sqrt{4\pi}$ and the allowed $m_0$ is bounded as mentioned in the previous section. Our investigation shows that successful leptogenesis explained by pure RG effects requires a lower range of the RHN mass scale $M_0$ than other effects discussed previously, which prefer $M_0\geq \mathcal{O}(10^{13})$  GeV~\cite{Adhikary:2008au,Karmakar:2014dva}.  Therefore,   the scale $M_0$ may be a clue to understanding  which source among RG, NLO, and softterm broken $A_4$ successfully generates the BAU data.

Recent investigation of the SS models that can generate consistent BAU data suggest that the  RHN neutrino mass scale  prefers  the range  below $\mathcal{O}(10^{8})$ GeV~\cite{Brdar:2019iem,Brivio:2019hrj}. The RHN scale is very interesting information to confirm which are the dominant sources generating consistent BAU data.
%%%%%%%%%%%%%%%%%%%%%%%%%%%%%%%%%%%%%%%%%%%%%%%%%%%%%%%%%%%%%%%%%%%%%%%%%%
\section{ \label{sec_cLFV} Lepton flavor violating decays $e_b\rightarrow e_a\gamma$}
In this section we study the effects of the allowed regions of parameter space satisfying leptogenesis on LFV decays.  Neutrino mixing is the only source of LFV processes.  The left- and right-handed bases of the original neutral neutrinos are  denoted as $\nu'_L=(\nu_L,\ (N_R)^c)^{\textrm{T}}$, where $\nu_L=(\nu_{1L},\ \nu_{2L},\ \nu_{3L})^{\textrm{T}}$ and $(N_R)^c=((N_{1R})^c,\; (N_{2R})^c,\ (N_{3R})^c)^{\textrm{T}}$. Also, we have $\nu'_R=(\nu'_L)^c=((\nu_L)^c,\ N_R)^{\textrm{T}}$. A  four-component spinor for a Majorana neutrino is then $\psi=(\psi_L,\ \psi_R)^{\textrm{T}}$, where $\psi=\nu_a,\ N_a$; $\psi_L=\nu_{aL},\ (N_{aR})^c$; and $\psi_R=(\nu_{aL})^c,\ N_{aR}$.  They satisfy $\psi^c=C\overline{\ps}^{\textrm{T}}=\psi$. The relations between a Majorana neutrino and the left- and right-handed components are $\psi_{L,R}=P_{L,R}\psi$, where $P_{L,R}=(1\mp\gamma_5)/2$. The total mass matrix of the neutrino is
\be M^{\nu}= \left(\begin{array}{cc}
	0  &  m_D\\
	m_D^T & M_R\\
\end{array}\right),\label{tnumass}\ee
where $m_D$ and $M_R$ are given in Eqs. (\ref{Majoranamass1}) and (\ref{Majoranamass2}), respectively. The  Lagrangian part describing the neutrino mass term is $-\frac{1}{2} \overline{\nu'_L}M^\nu (\nu_L')^c + \mathrm{H.c.}$ The mass matrix in Eq. (\ref{tnumass}) is diagonalized by the mixing matrix  $U^{\nu}$, which is unitary and  satisfies
\be U^{\nu T}M^{\nu} U^{\nu}=\hat{M}^{\nu}=\mathrm{diag}(m_{n_1},\;m_{n_2},\dots,\;m_{n_6})\simeq \mathrm{diag}(m_1,\,m_2,\,m_3,\,M_1,\,M_2,\,M_3),   \label{diamnu}\ee
where the first three mass values $m_{n_a}$ ($a=1,2,3$) and respective eigenvectors $n_{a}$ are identified with those of active neutrinos observed by experiments.  The remaining masses belong to three heavy neutrinos $n_{4,5,6}$. Hence, the last term in Eq. (\ref{diamnu}) is derived from the relations shown in Eqs. \eqref{eq_lightactivenu} and  \eqref{D_MR}. The relations between the original and mass basis of the neutrino are
\be \nu'_{iL}= U^{\nu*}_{ij}n_{jL}=U^{\nu*}_{ij} P_L n_j,\hs \; \nu'_{iR}= U^{\nu}_{ij} n_{jR}=U^{\nu}_{ij} P_R n_j,  \label{nurelate}\ee
where $n=(n_1,\; n_2,\; \dots,\; n_6)^{\textrm{T}}$ and $n_i=(n_{iL},\;n_{iR})^{\textrm{T}}$ ($i=1,2,\dots,6$).

Based on previous parmeterizations~\cite{Casas:2001sr,Ibarra:2010xw}, the matrix $U^{\nu}$ can be written as
\bea  U^{\nu}=\left(
\begin{array}{cc}
	I_3 & \textbf{O}\\
	\textbf{O}&U_R  \\
\end{array}
\right) \exp\left(
\begin{array}{cc}
	\textbf{O} &  R\\
	-R^{\dagger}&\textbf{O} \\
\end{array}
\right)\left(
\begin{array}{cc}
	U_{\mathrm{PMNS}} & \textbf{O}\\
	\textbf{O}&V_3  \\
\end{array}
\right),
\label{Unugen}\eea
where $\textbf{O}$ is the $3\times3$ matrix with all elements being zeros;   $U_R$, $V_3$,  and $U_{\mathrm{PMNS}}$ are three $3\times3$ unitary matrices;  and $R$ is a $3\times 3$ matrix satisfying $|R|\equiv \mathrm{max}[|R_{ij}|]\ll1$  for all $i,j=1,2,3$.  Apart from Eq.~\eqref{active mass}, other  SS relations for determining $R$ and heavy neutrino masses are identified up to $\mathcal{O}(R^2)$ as follows
\begin{align}
	\label{eq_SSrelation}
	R^*&= \left( m_DU_R\right) \left[U_R^{\textrm{T}}M_{N}U_R\right]^{-1},\crn
	V^*_3 \hat{M}_RV_3&= U_R^{\textrm{T}}M_{N}U_R +\frac{1}{2}R^{\textrm{T}}R^*U_R^{\textrm{T}}M_{N}U_R +\frac{1}{2}U_R^{\textrm{T}}M_{N}U_RR^{\dagger}R,
\end{align}
where we have applied the result from Refs.~\cite{Casas:2001sr,Ibarra:2010xw},  after taking a rotation of $M^{\nu}$ corresponding to the first matrix in the right-hand side of Eq.~\eqref{Unugen}, which gives $m_D\rightarrow m_DU_R$ and $M_R\rightarrow U^{\textrm{T}}_RM_RU_R$.

In our framework,  $U_R$ is defined from Eq.~\eqref{D_MR},  and $U_{\mathrm{PMNS}}=U_R^*$ is the well-known mixing matrix of active neutrinos defined in Eq.~\eqref{eq_UPMNS0}. Therefore, it can be proved that
\begin{align}
	R^*&=\sqrt{\frac{m_0}{M_0}}U^*_{\mathrm{PMNS}}\times \mathrm{diag}\left(\frac{M_0}{M_1},\,\frac{M_0}{M_2},\,\frac{M_0}{M_3}\right), \crn
	V^*_3 \hat{M}_RV_3&= \mathrm{diag}(M_1,\,M_2,\,M_3) +\frac{m_0}{2}\times \mathrm{diag}\left(\frac{M_0}{M_1},\,\frac{M_0}{M_2},\,\frac{M_0}{M_3}\right)U_{\mathrm{PMNS}}U^*_{\mathrm{PMNS}} \crn
	&+\frac{m_0}{2}\times U^*_{\mathrm{PMNS}}U_{\mathrm{PMNS}}\mathrm{diag}\left(\frac{M_0}{M_1},\,\frac{M_0}{M_2},\,\frac{M_0}{M_3}\right). \label{eq_SSrelation1}
\end{align}
In the allowed region we have  $\mathcal{O}(10^{-11}\ \mathrm{GeV})\sim m_0\ll M_{1,2,3}$, $ M_{1,2,3}\geq \mathcal{O}(10^6) \ \mathrm{GeV}$, and $M_0/M_{1,2,3}=\mathcal{O}(1)$, and hence the  assumption mentioned above that heavy neutrino masses are given by Eq.~\eqref{D_MR} is acceptable with a very high accuracy. Using this approximation we also get $V_3=I_3$.

Up to the order $\mathcal{O}(R^2)$, the mixing matrix $U^{\nu}$  is now
\bea  U^{\nu}\simeq\left(
\begin{array}{cc}
	(1-\frac{ 1}{2}RR^{\dagger})U_{\mathrm{PMNS}} &  R\\
	-U^*_{\mathrm{PMNS}}R^{\dagger}U_{\mathrm{PMNS}}&U^*_{\mathrm{PMNS}}\left(1-\frac{ 1}{2}R^{\dagger}R\right)  \\
\end{array}
\right)+ \mathcal{O}(R^3).
\label{Unu1}\eea
Finally, $U^{\nu}$ can be presented as a function  of $M_0,\;m_0,\  \phi$ and $\kappa$.
As we know, in the minimal  model (MSS), where only heavy Dirac neutrinos are added in the SM to explain the neutrino oscillation data, the branching ratio (Br) of the LFV decay Br$(e_b\rightarrow e_a\gamma)$ was shown to be suppressed, for example Br$(\mu \rightarrow e\gamma)\leq \mathcal{O}(10^{-54})$. On the other hand,  some SM extensions with heavy neutrinos obeying the SS mechanism~\cite{Minkowski:1977sc,Mohapatra:1979ia,GellMann:1980vs,Yanagida:1979as,Schechter:1980gr} can give large Br$(e_b\rightarrow e_a\gamma)$, close to the recent experimental sensitivities~\cite{Gorbunov:2014ypa}. In our model, the presence of the charged Higgs boson gives another one-loop contribution to the LFV decay amplitude. This leads to a different prediction for LFV decays that deserves to be investigated. It should be noted that, although in the model under consideration the properties of the charged Higgs boson may be the same as those discussed thoroughly in Refs.~\cite{Branco:2011iw,Hung:2019jue}, the LFV couplings with neutrinos particularly, the behaviors of the LFV may be more predictive than the results of an LFV investigation for 2HDM discussed recently in Ref.~\cite{Vicente:2019ykr}.

\begin{figure}[h]
	\centering
	% Requires \usepackage{graphicx}
	\includegraphics[width=12cm]{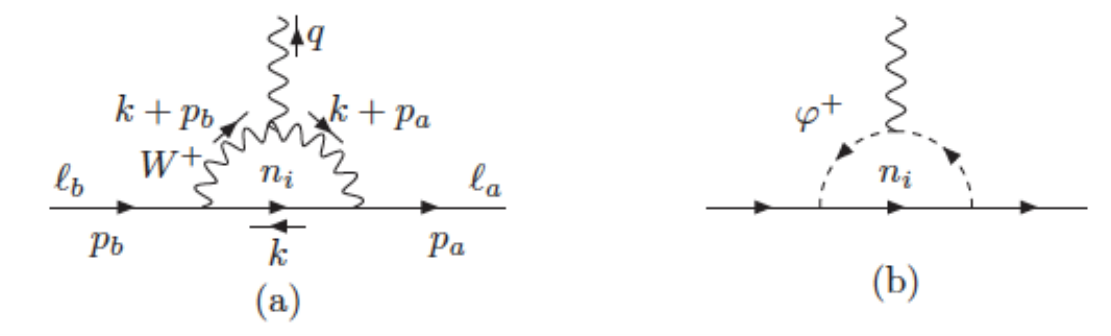}\\
	\caption{One-loop Feynman diagrams contributing to $C_{L,R}$ for the decay $e_b\rightarrow e_a\gamma$ in the unitary gauge.}\label{liljga1}
\end{figure}

In the Yukawa Lagrangian part of Eq. (\ref{Lpartlep}), couplings relating to LFV decays are
\bea -\mathcal{L} &\rightarrow& \frac{m_{e_a}}{\sqrt{2}v_d} S_d \overline{e_a}e_a+ \frac{m_{e_a}}{v_d}\left(\overline{\nu_{La}}e_{Ra}H^+_d+ \mathrm{H.c.}\right) \crn
&&+ \frac{S_u}{\sqrt{2}}p\left(\overline{\nu_{aL}}N_{Ra}+\mathrm{h.c.}\right)  +p\left[ H^-_u \overline{e_{aL}}N_{Ra} + \mathrm{H.c.} \right]. \label{LFVcoup}\eea
Using the transformations to the physical states of the charged Higgs boson and neutrinos given respectively in Eqs.~\eqref{nurelate} and  \eqref{cHiggs}, the Feynman rules for LFV couplings of the charged Higgs boson are collected in Table~\ref{coupling}. The Feynman rules for LFV coupling relating with the $W$ boson using our notation can be found in Ref.~\cite{Thao:2017qtn}. They are consistent with those mentioned in 2HDMs~\cite{Branco:2011iw}. All the Feynman rules for calculating  amplitudes of LFV decays $e_b \rightarrow e_a\gamma$ in the unitary gauge are shown in Table~\ref{coupling}. Accordingly, the one-loop calculations in this work will be done in the unitary gauge.
\begin{table}[ht]
	\centering
	\caption{ Couplings relating with one-loop three-point Feynman diagrams that contribute to the LFV decay $e_i\rightarrow e_j\gamma$ in the unitary gauge. }
	\begin{tabular}{|c|c|c|c|}
		\hline
		Vertex & Coupling & Vertex & Coupling \\
		\hline
		$\overline{e_a}n_i\varphi^-$ &  $\dfrac{-igU^{\nu*}_{ai}}{m_W\sqrt{2}}\left(m_{e_a}t_{\beta} P_L+m_{n_i}t^{-1}_{\beta} P_R \right)$&$\overline{n_i}e_a\varphi^+$ &  $\dfrac{-igU^{\nu}_{ai}}{m_W\sqrt{2}}\left(m_{e_a}t_{\beta} P_R+m_{n_i}t^{-1}_{\beta} P_L \right)$ \\
		\hline
		$\overline{e_a}n_iW^-_\mu$&$\dfrac{ig}{\sqrt{2}}U^{\nu*}_{ai}\gamma^\mu P_L$&$\overline{n_i}e_aW^+_\mu$&$\dfrac{ig}{\sqrt{2}} U^{\nu}_{ai}\gamma^\mu P_L$\\
		\hline
	\end{tabular}\label{coupling}
\end{table}
The Br of the LFV decays $e_b\rightarrow e_a\gamma$ ($m_{e_b}>m_{e_a}$), where $(e_b,~e_a)=$ $\{(\tau,\mu),$ $(\tau,e),$ $(\mu,~e)\}$, can be determined  as follows~\cite{Lavoura:2003xp}:
\begin{equation}\mathrm{Br}(e_b\rightarrow e_a\gamma)= \left(1-\frac{m_a^2}{m_b^2}\right)^3 \times \frac{12\pi^2}{G_F^2m_b^2}\left(|C_L|^2+ |C_R|^2\right)\times \mathrm{Br}(e_b\rightarrow e_a\bar{\nu}_{a}\nu_b), \label{brlfvdecay1}
\end{equation}
where $C_{L,R}$ are scalar factors arising from loop corrections. In the unitary gauge, one-loop  Feynman diagrams contributing to $C_{L,R}$ are shown in Fig.~\ref{liljga1}.

The Br of the decay can therefore be calculated through the well-known decay rates  $\tau$ or $\mu$, namely Br$(e_b\rightarrow e_a\bar{\nu}_{a}\nu_{b})$.  The corresponding partial decay width is $\Gamma (e_b\rightarrow e_a\bar{\nu}_{a}\nu_{b})=   \frac{G_{\textrm{F}}^2m^5_{b}}{192\pi^3}$
with $G_{\textrm{F}}=\frac{g^2\sqrt{2}}{8m^2_W}$. The  experimental values are  $\mathrm{Br}(\tau\rightarrow\mu\bar{\nu}_{\mu}\nu_{\tau}) \simeq 17.41\%$,  $\mathrm{Br}(\tau\rightarrow e\bar{\nu}_{e}\nu_{\tau}) \simeq 17.83\%$, and  $\mathrm{Br}(\mu\rightarrow e\bar{\nu}_{e}\nu_{\mu}) \simeq 100\%$.

In the limit of zero external momenta $m_W^2, m^2_{\varphi} \gg p_a^2,p_b^2\rightarrow 0$, the analytic expressions of the amplitude are
\be  C_{L,R}= C^{W}_{L,R}+C^{\varphi}_{L,R},
\label{CLR}\ee
where  $C^{W}_{L,R}$ and $C^{\varphi}_{L,R}$ are determined in Appendix \ref{CLR1},  consistent with Ref. \cite{Lavoura:2003xp}.
%---
For low energy, Eq.~\eqref{brlfvdecay1} can be written in a more convenient form  as
\begin{equation}\mathrm{Br}(e_b\rightarrow e_a\gamma)= \left(1- \frac{m_a^2}{m_b^2}\right)^3 \times \frac{3\alpha_{\mathrm{e}}}{2\pi}\left( \frac{m_a^2}{m_b^2}\left| D_L \right|^2+ |D_R|^2\right)\times \mathrm{Br}(e_b\rightarrow\,  e_a\bar{\nu}_{a}\nu_b), \label{brlfvdecay2}
\end{equation}
where $C_{L,R}=\frac{g^2e m_{a,b}}{32\pi^2 m_W^2}\times D_{L,R}$ and $\alpha_e\equiv e^2/(4\pi)\simeq 1/137$  in numerical investigations.

\begin{figure}[ht]
%fig_eijgaNH	
	\minipage{0.485\textwidth}
\centering
\includegraphics[width=7.5cm]{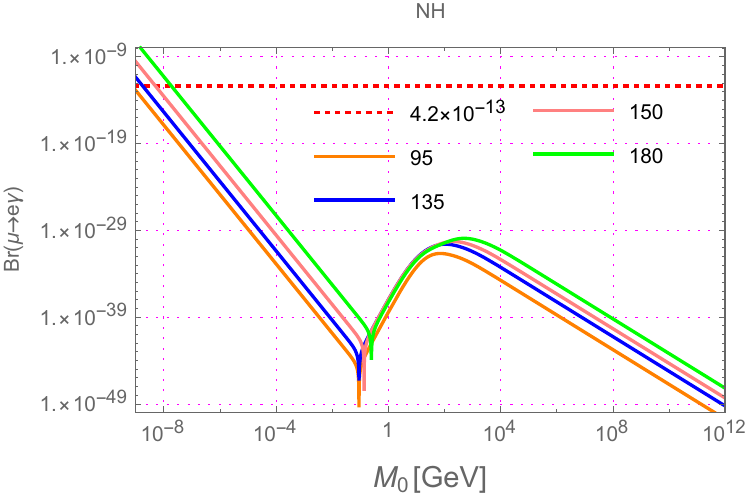}
\endminipage
\hfill
\quad
\minipage{0.485\textwidth}
\centering
\includegraphics[width=7.5cm]{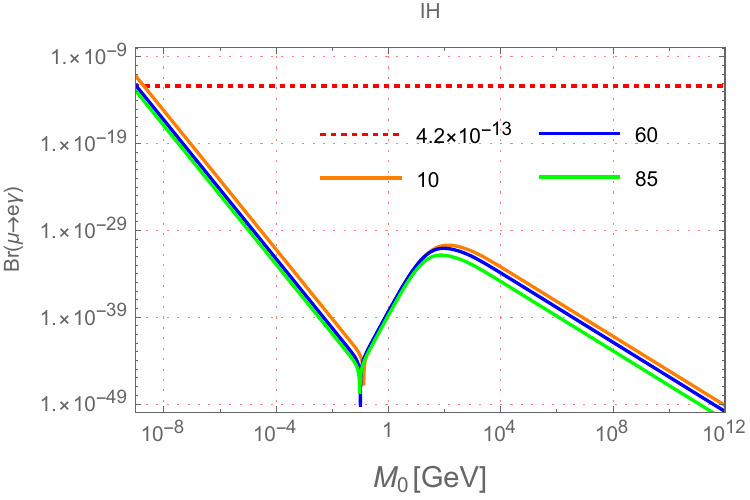}
\endminipage
\hfill
	\caption{Br$(\mu\rightarrow\,e\gamma)$ as a function of $M_0$ with different allowed values of $(\phi,\kappa)$ given in Table~\ref{table_KappaPhi} and \ref{table_KappaPhiI} for the NH and IH cases, where the values of $\phi=95^\textrm{o},\; 135^\textrm{o},\; 150^\textrm{o},\; 180^\textrm{o}$ or $10^\textrm{o},\ 60^\textrm{o},\ 85^\textrm{o}$ are pointed out in the respective figures.  The red dotted lines show the experimental upper bound Br$(\mu \rightarrow \,e \gamma)<4.2\times 10^{-13}$. } \label{fig_cLFV}
\end{figure}

Because the charged Higgs boson have similar properties to those given in the 2HDM, we set a lower bound of $300\ \mathrm{ GeV}\leq m_{\varphi}\leq 2000\ \mathrm{ GeV}$, and $0.3\leq t_{\beta}\leq 10$. Our investigation shows that the qualitative results of LFV decay  do not change significantly, hence in the following illustration we fix $t_{\beta}=3$ and $m_{\varphi}=500$ GeV.  With different pairs of $(\phi, \kappa)$ given in Table~\ref{table_KappaPhi} satisfying all experimental data of neutrino oscillation, the dependence of LFV decay Br$(\mu\rightarrow\,e\gamma)$ on the heavy RHN mass scale $M_0$ is shown in Fig.~\ref{fig_cLFV}. 
We constrain the lower bound of exotic neutrino masses by $M_0\ge \mathcal{O}(1)$ eV, leading to $|R|\sim \sqrt{\frac{m_0}{M_0}}\leq \mathcal{O}(10^{-1})$, so that the seesaw mechanism still works well.

\begin{figure}[ht]
		\minipage{0.485\textwidth}
	\centering
	\includegraphics[width=7.5cm]{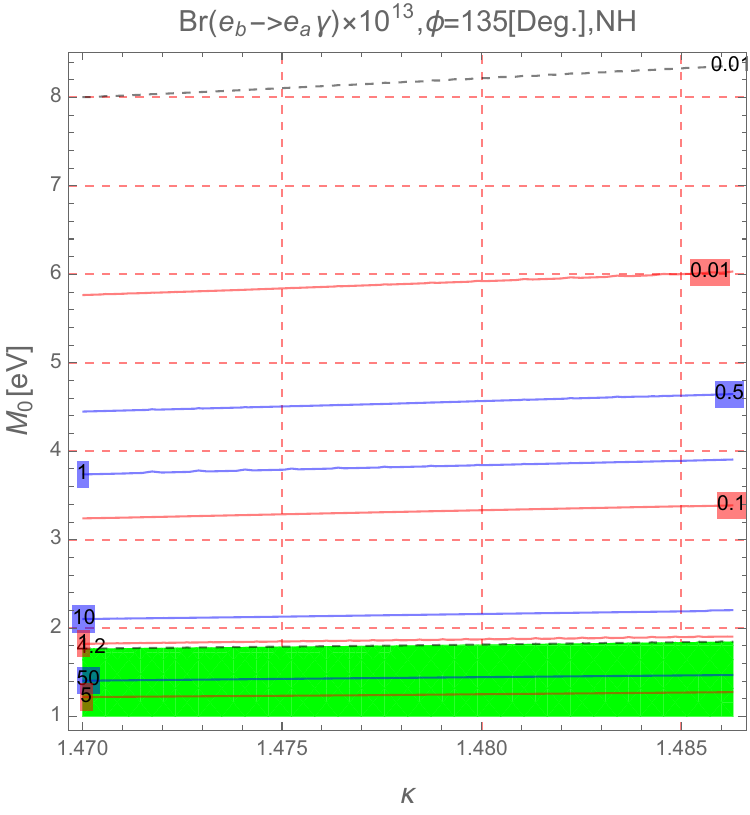}
	\endminipage
	\hfill
	\quad
	\minipage{0.485\textwidth}
	\centering
	\includegraphics[width=7.5cm]{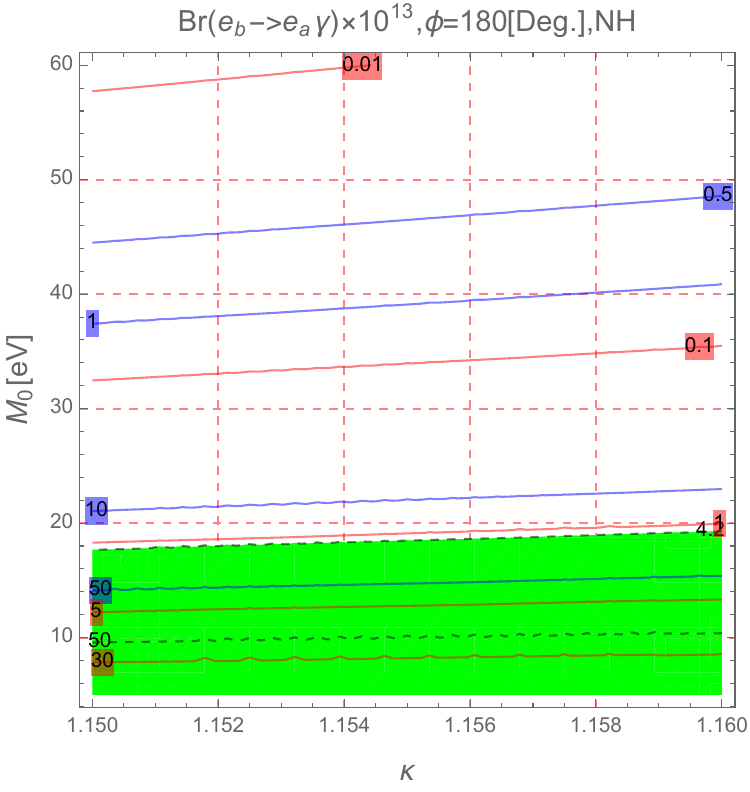}
	\endminipage
	\hfill
	\caption{Contour plots of Br$(e_b \rightarrow\,e_a\gamma)$ as functions of $\kappa$ and $M_0$ in the NH case. The blue regions are excluded by Br$(\mu \rightarrow\, e\gamma)<4.2\times 10^{-13}$. The black dashed, red and blue curves show the constant values of Br$(\mu\rightarrow e\gamma)$, Br$(\tau\rightarrow e\gamma)$, and Br$(\tau\rightarrow \mu \gamma)$, respectively.}\label{fig_eijgacontour}
\end{figure}

\begin{figure}[ht]
	\minipage{0.485\textwidth}
	\centering
	\includegraphics[width=7.5cm]{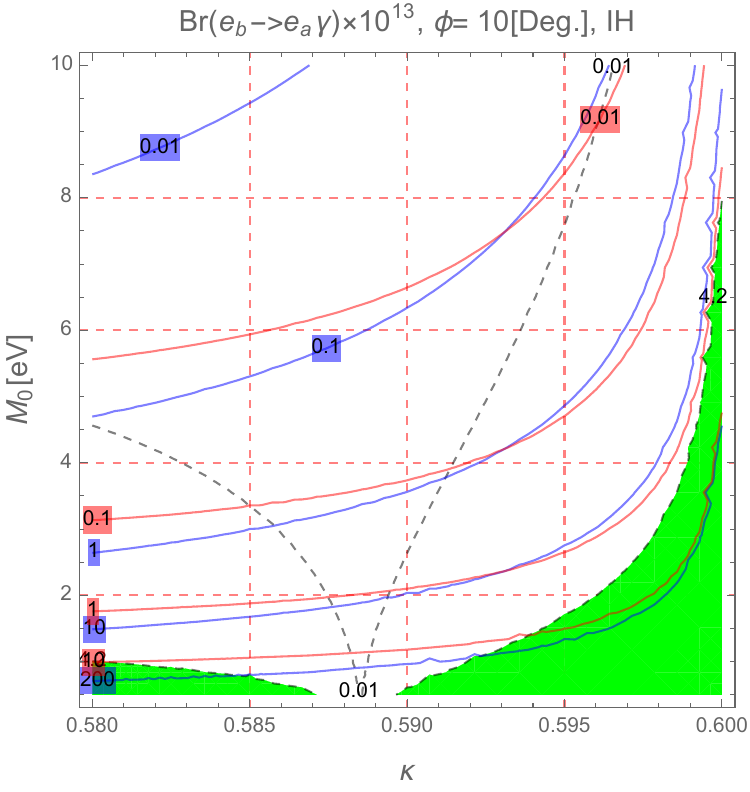}
	\endminipage
	\hfill
	\quad
	\minipage{0.485\textwidth}
	\centering
	\includegraphics[width=7.5cm]{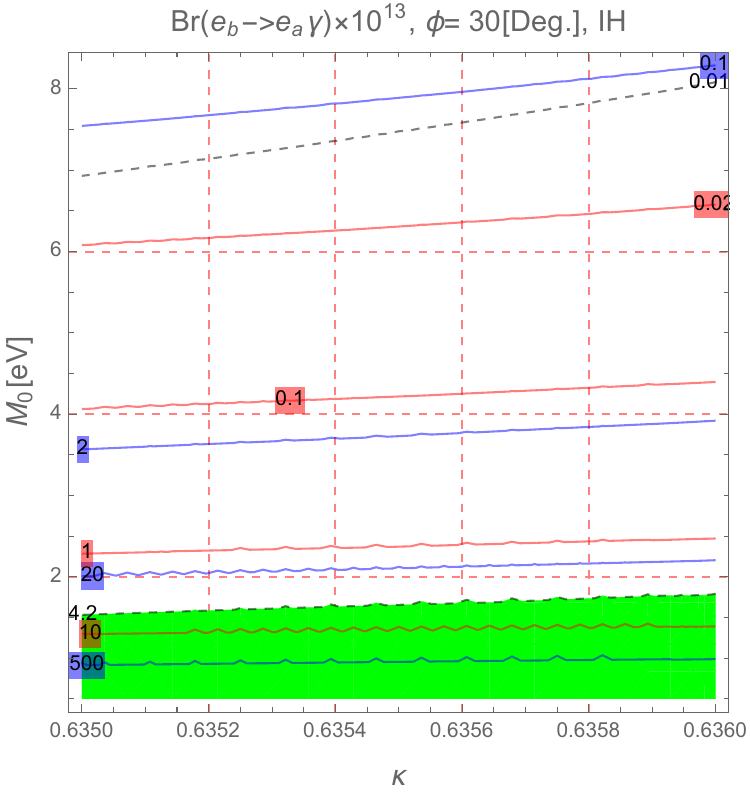}
	\endminipage
	\hfill
	\caption{Contour plots of Br$(e_b \rightarrow\,e_a\gamma)$ as functions of $\kappa$ and $M_0$ in the NH case. The blue regions are excluded by Br$(\mu \rightarrow\, e\gamma)<4.2\times 10^{-13}$. The black dashed, red and blue curves show the constant values of Br$(\mu\rightarrow e\gamma)$, Br$(\tau\rightarrow e\gamma)$, and Br$(\tau\rightarrow \mu \gamma)$, respectively.}\label{fig_eijgacontourI}
\end{figure}

The point to note is that Br$(\mu\rightarrow\,e\gamma)$ can approach the current experimental sensitivity Br$(\mu \rightarrow\,e\gamma)<4.2\times 10^{-13}$~\cite{TheMEG:2016wtm} in the light exotic mass region, namely $M_{1,2,3}= \mathcal{O}(M_0)\sim 10^{-9}-10^{-8}$ GeV.  On the other hand,  Br$(\mu\rightarrow\,e\gamma)$ is very suppressed with heavy $M_i$. Hence, if cLEV decays are detected,  the region of heavy exotic neutrino masses is excluded, implying that leptogenesis and LFV data cannot be explained simultaneously in  the model under consideration, i.e. the model is ruled out.   We can see that the allowed region of Br$(\mu\rightarrow\,e\gamma)<4.2\times 10^{-13}$ results in very small values of  Br$(\tau\rightarrow\,e\gamma)$ and Br$(\tau\rightarrow\,\mu\gamma)$; see an illustration in Figs.~\ref{fig_eijgacontour} and ~\ref{fig_eijgacontourI} for the HN and IH cases, respectively.

Generally,  the constraint Br$(\mu\rightarrow e\gamma)<4.2\times 10^{-13}$ results in Br$(\tau\rightarrow \mu \gamma)\leq \mathcal{O}(10^{-12})$ and Br$(\tau\rightarrow e \gamma)\leq \mathcal{O}(10^{-13})$. These values are still much smaller than the sensitivity of near-future experiments~\cite{TheMEG:2016wtm,Baldini:2013ke,Baldini:2019elc,Aubert:2009ag,Aushev:2010bq}.
In most allowed regions of parameters obtained from neutrino oscillation data, all of the values of Br$(e_b\rightarrow\,e_a\gamma)$ satisfy the experimental data of LFV decay, including in the region with heavy enough RHN masses  to successful explain the leptogenesis data.

\begin{figure}[ht]
	\minipage{0.485\textwidth}
	\centering
	\includegraphics[width=7.5cm]{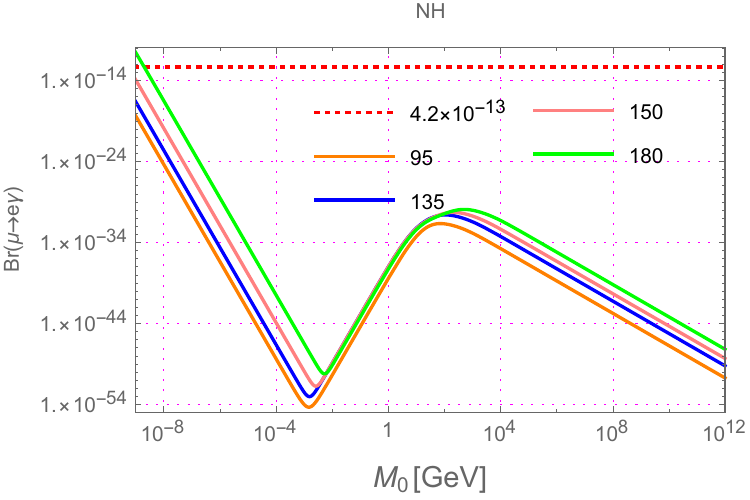}
	\endminipage
	\hfill
	\quad
	\minipage{0.485\textwidth}
	\centering
	\includegraphics[width=7.5cm]{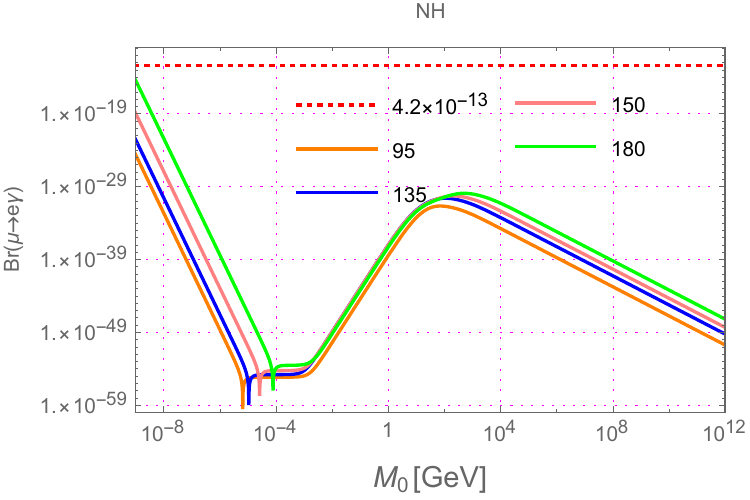}
	\endminipage
	\hfill
	\caption{Plots of Br$(e_b \rightarrow\,e_a\gamma)$ as functions of $\kappa$ and $M_0$ in the NH case, where $U^{\nu}$ is kept up to $\mathcal{O}(R^4)$ and $\mathcal{O}(R^6)$ in the left and  right panels, respectively.}\label{fig_eijgak}
\end{figure}

In the above discussion, the neutrino mixing matrix defined in Eq.~\eqref{Unugen} is normally kept up to the order of $\mathcal{O}(R^2)$. The results may not be accurate for very light $M_0$; see the illustrations for the NH case shown in Fig.~\ref{fig_eijgak}, where higher orders of $R$ are included.
Anyway, the Br$(\mu\rightarrow\,e\gamma)$, Br$(\tau\rightarrow\,e\gamma)$, and Br$(\tau\rightarrow\,\mu\gamma)$ are always well below the current experimental upper bounds.  We note that our results predict that Br$(\mu\rightarrow e\gamma)<\mathcal{O}(10^{-29})$  with $1 \,\mathrm{GeV}<M_0<10^4$ GeV.  This is different from the results discussed in some previous work showing  that the Br$(\mu\rightarrow\,e\gamma)$ can reach the current experimental bound~\cite{Ibarra:2011xn}.  The reason is that the Dirac matrix mass in Ref.~\cite{Ibarra:2011xn} is  defined following the Casas$-$Ibarra parameterization~\cite{Casas:2001sr}.

\section{Conclusion}
We have studied the seesaw version of an $A_4$ flavor symmetry model with two Higgs singlets beside other scalars as usual $A_4$ models. The allowed regions of the parameter space satisfying the current experimental neutrino oscillation data at 3 $\sigma$ CL. are given numerically. We have found that the allowed ranges of $\kappa$ and $\phi$  corresponding to the NH and the IH schemes separate completely. In particular,  $\phi \in (90^{\textrm{o}},270^{\textrm{o}})$ and $\kappa \in(1.15,1.5)$ are allowed for the NH case, while $\phi \in (0^{\textrm{o}},90^{\textrm{o}})\cup(270^{\textrm{o}},360^{\textrm{o}})$ and $\kappa \in(0.55,1)$ are allowed for the IH case. The model then predicts that the possible values of $|\langle m \rangle|$ will be 0.002 eV $\leq |\langle m\rangle| \leq$ 0.038 eV for the NH and 0.048 eV $\leq |\langle m\rangle| \leq$ 0.058 eV for the IH. This prediction is testable by running $0\nu 2\beta$ decay experiments, therefore is very clear information to confirm which NH or IH scheme is realistic. We have shown that the diagonal Hermitian matrix $H = Y_\nu Y_\nu^\dag$ in the original model  becomes non-diagonal  after the effect of renormalization group evolution is included, therefore  leptogenesis can be generated successfully in the allowed regions.  The RHN mass scale $M_0 = 10^8-10^{12}$ GeV is required  for successful leptogenesis. In this range, it  decreases with higher values of $\tan\beta$. Illustrations for $\eta_B$ as functions of  $m_0$, $\phi$,  $|\langle m\rangle|$, and $M_0$ for different $t_{\beta}$ have been presented. The minimum value of $M_0$ ($10^8$ GeV) corresponds to the so-called resonant leptogenesis where two heavy RHN masses $M_1$ and $M_3$ are almost degenerate (and also corresponds to the maximum value of $|\langle m\rangle|$ predicted by the model).  We  have found an interesting correlation between $\eta_B$  and  $|\langle m \rangle|$, so that once $|\langle m \rangle|$ is confirmed, we can pin down the RHN masses for successful leptogenesis for some given values of $\tan\beta$ as well as the absolute values of active neutrino masses.

We have also investigated the LFV decays of charged leptons, $e_a \rightarrow e_b\gamma$. Our investigation shows that if this signal is found experimentally in the future, the RHN mass scale  must be smaller than the order of $\mathcal{O}(10\ \mathrm{eV})$, so that the class of  models we mentioned above must be improved to explain both LFV decays and leptogenesis, or they will be ruled out.

%%%%%%%%%%%%%%%%%%%%%%%%%%%%%%%%%%%%%%%%%

\section*{Acknowledgments}
This research is funded by Vietnam  National Foundation for Science and Technology Development (NAFOSTED) under grant number 103.01-2018.331.
\appendix
\newpage
\section{\label{A4rules}$A_4$ group: the AF (Altarelli $-$ Feruglio) basis introduced by  G. Altarelli and F. Feruglio}
%\subsection{In the AF (Altarelli-Feruglio) basis introduced by  G. Altarelli and F. %Feruglio}
The non-Abelian $A_4$ is a group of even permutations of four objects and has $4!/2=12$ elements. The group is generated by two generators $S$ and $T$ satisfying the relations
\begin{eqnarray}
S^2 = (ST)^3 = (T^3) = 1.
\end{eqnarray}
There are three one-dimensional irreducible representations of the group denoted as
\bea
1: && S  = 1, \hs T = 1,\\
1': && S  = 1, \hs T =  e^{i4\pi/3}\equiv \om^2,\\
1'':&& S  = 1, \hs T = e^{i2\pi/3}\equiv \om.
\eea
It is easy to check that there is no two-dimensional irreducible representation of this group. The three-dimensional unitary representations of $T$ and $S$ are given by
\bea
T =\left(\begin{array}{ccc}
	1  & 0  &  0\\
	0  &   \om^2  &  0 \\
	0  &   0  & \om \end{array}\right),\hs
S= \frac{1}{3}{\left(\begin{array}{ccc}
		-1 & 2 & 2\\
		2 & -1 & 2 \\
		2 & 2  & -1\end{array}\right)},\label{AFbasis}
\eea
where $T$ has been chosen to be diagonal. The multiplication rules for the singlet and triplet representations corresponding to the above basis of two generators $T, S$ are given as
\be
1\times1 = 1, \ 1'\times1'' = 1,\ 3\times3 = 3 + 3_A +1 + 1' + 1''.
\ee
For triplets
\be
a = (a_1, \ a_2, \ a_3), \hs b=(b_1, \ b_2,\ b_3),
\ee
one can write
\bea
1 \equiv (ab) &=& (a_1b_1+a_2b_3+a_3b_2),\\
1' \equiv (ab)' &=& (a_3b_3+a_1b_2+a_2b_1),\\
1'' \equiv (ab)'' &=& (a_2b_2+a_1b_3+a_3b_1).
\eea
Note that while 1 remains invariant under the exchange of the second and the third elements of $a$ and $b$, $1'$ is symmetric under the exchange of the first and second elements while $1''$ is symmetric under the exchange of the first and third elements.
\bea
3 &\equiv & (ab)_S \nn\\
  &=& \frac{1}{3}(2a_1b_1-a_2b_3-a_3b_2, 2a_3b_3-a_1b_2-a_2b_1, 2a_2b_2-a_1b_3-a_3b_1),\\
3_A &\equiv & (ab)_A ~=~\frac{1}{2}(a_2b_3-a_3b_2, a_1b_2-a_2b_1, a_3b_1-a_1b_3).
\eea

We will only focus only on 3 since the $3_A$ terms are antisymmetric and hence cannot be used for the neutrino mass matrix. In the triplet 3, we can see that the first element has 2$-$3 exchange symmetry, the second element has 1$-$2 exchange symmetry, while the third element earns 1$-$3 interchange symmetry.

Moreover, if $c, c', c''$ are singlets of  type $1, 1',  1''$, and $a = (a_1, \ a_2, \ a_3)$ is a triplet, then the products $ac, ac', ac''$ are triplets explicitly given by $(a_1c,\ a_2c,\ a_3c)$, $ (a_3c',\ a_1c',\ a_2c')$, $(a_2c'',\ a_3c'',\ a_1c'')$, respectively.

Because the above basis, $T$ is complex and $T^{*}\neq T$ in general, the complex conjugate representation $r^*$ of a representation $r$ ($r=1',\ 1'',\ 3$) is not the same as $r$. It is determined by the following rules \cite{Feruglio:2008ht,Feruglio:2009hu}:
\bea c \sim 1\rightarrow c^* \sim 1 ,\;&& c' \sim 1' \rightarrow {c'}^* \sim  {1'}^*=1'', \;  c' \sim 1' \rightarrow {c''}^* \sim  {1''}^*=1',\crn
a=(a_1,\ a_2,\ a_3)  &\sim& 3 \rightarrow a^*=(a^*_1,\ a_3^*,\ a_2^*).  \label{a4crules}\eea
For the one-dimensional reps, it is easy to see these properties because $(\omega^2)^*=\omega$.  For the 3-reps we can find a transformation $U$ that changes $3^*$ into $3$ or $3^*\sim3$ and vice versa. This is similar to the case of $SU(2)$ symmetry. Namely,   $UTU^{-1}=T^*=T^2$ and $USU^{-1}=S^*=S$ for $T$ and $S$ given in Eq.~\eqref{AFbasis}. We can see this  in the $S_4$ group where all of $T$, $T^2$, and $S$ are in the same conjugate class; see the details in Refs.~\cite{Ishimori:2010au, Bazzocchi:2009pv}. Hence,  $U$ belongs to $S_4$ but not $A_4$, namely
\begin{equation}\label{eq_Utts}
U=TSTS^2=\left(
\begin{array}{ccc}
1 & 0 & 0 \\
0 & 0 & 1 \\
0 & 1 & 0 \\
\end{array}
\right).
\end{equation} 
In the model considered, the $A_4$ lepton triplet $\overline{\psi^l}=(\overline{\psi^l_1},\;\overline{\psi^l_2},\;\overline{\psi^l_3})\sim 3$ has a complex conjugate of $\psi^l=(\psi^l_1,\;\psi^l_3,\;\psi^l_2)\sim3^*$. The $3\times 3^*$ is used for constructing the kinetic terms of lepton and Higgses, the  Higgs potential,\dots For example some quadratic terms respecting $A_4$ symmetry  are:
\bea \overline{\psi^l}&=&(\overline{\psi^l_1},\;\overline{\psi^l_2},\;\overline{\psi^l_3})\sim 3,\crn
\rightarrow && \left(\overline{\psi^l}\ga^{\mu}D_{\mu}\psi^l\right)_{1}= \overline{\psi^l_1}\ga^{\mu}D_{\mu}\psi^l_1+ \overline{\psi^l_2}\ga^{\mu}D_{\mu}\psi^l_2+ \overline{\psi^l_3}\ga^{\mu}D_{\mu}\psi^l_3,\crn
\phi_S&=&(\phi_{S_1},\; \phi_{S_2},\;\phi_{S_3})\sim 3, \;\phi^*_S=(\phi^*_{S_1},\; \phi^*_{S_3},\;\phi^*_{S_2})\sim 3^*\crn
\rightarrow && \left((D^\mu\phi_S)^{\dagger}D_{\mu}\phi_S\right)_{1}=(D^\mu\phi_{S_1})^{\dagger}D_{\mu}\phi_{S_1} +(D^\mu\phi_{S_2})^{\dagger}D_{\mu}\phi_{S_2} +(D^\mu\phi_{S_3})^{\dagger}D_{\mu}\phi_{S_3}, \crn
&& \left((D^\mu\phi_T)^{\dag}D_{\mu}\phi_T\right)_{1}=(D^\mu\phi_{T_1})^{\dagger}D_{\mu}\phi_{T_1} +(D^\mu\phi_{T_2})^{\dag}D_{\mu}\phi_{T_2} +(D^\mu\phi_{T_3})^{\dag}D_{\mu}\phi_{T_3}, \crn
\xi'&\sim& 1'\rightarrow {\xi'}^*\sim {1'}^*=1''\rightarrow ({\xi'}^*\xi')_1={\xi'}^*\xi', \hs ({\xi''}^*\xi'')_1={\xi''}^*\xi''.
\label{kinetic}\eea
Note that the AF basis was used in Ref. \cite{King:2011zj}.
\section{\label{HiggsPotential} Higgs potential and vacuum stability}

Now we come to consider the Higgs potential which satisfies the condition of $A_4$ invariance,
\begin{align}
V_{H'}&=\mu_1^2 h_u^{\dag}h_u+\mu_2^2 h_d^{\dag}h_d+ \mu_3^2 \xi'^\dag\xi'
+ \mu_4^2 \xi''^\dag\xi''
\crn&+ \la_1  (h_u^{\dag}h_u)^2  + \la_2  (h_d^{\dag}h_d)^2 + \la_3 (h_u^{\dag}h_u)(h_d^{\dag}h_d)\crn
&+ \la_4 (h_u^{\dag}h_d) (h_d^{\dag}h_u)+ \la^{\xi'}(\xi'^\dag\xi')^2 + \la^{\xi''}(\xi''^\dag\xi'')^2+\la^{\xi'\xi''}(\xi^{'\ast}\xi^{'})(\xi^{''\ast}\xi^{''})\crn
&+\la^{u\xi'}(h_u^{\dag}h_u)(\xi'^\ast\xi')+ \la^{d\xi'}(h_d^{\dag}h_d)(\xi'^\ast\xi')+
\la^{u\xi''}(h_u^{\dag}h_u)(\xi''^\ast\xi'') \crn
&+\la^{d\xi''}(h_d^{\dag}h_d)(\xi''^\ast\xi'')\crn
&+V(\phi_T)+V(\phi_S)+V(\phi_T,\phi_S)+V(\phi_T, h_u) + V(\phi_T, h_d)  \crn
&+V(\phi_S, h_u)+V(\phi_S, h_d)  +V(\phi_T,\xi',\xi'')+V(\phi_S,\xi',\xi''),\label{Hpotential}
\end{align}
where
\begin{align}
V(\phi_T)&=\mu^2_T(\phi^\dag_T\phi_T)_1+\la^{\phi_T}_1(\phi^\dag_T\phi_T)_1(\phi^\dag_T\phi_T)_1 + \la^{\phi_T}_{2}(\phi^\dag_T\phi_T)_{1'}(\phi^\dag_T\phi_T)_{1''} \crn
& +  \la^{\phi_T}_{3}(\phi^\dag_T\phi_T)_{3_A}(\phi^\dag_T\phi_T)_{3_A}+\la^{\phi_T}_{4}(\phi^\dag_T\phi_T)_{3_S}(\phi^\dag_T\phi_T)_{3_S} \crn
&+ \la^{\phi_T}_{5}(\phi^\dag_T\phi_T)_{3_S}(\phi^\dag_T\phi_T)_{3_A},\crn
V(\phi_S)=&\mu^2_S(\phi^\dag_S\phi_S)_1+\la^{\phi_S}_1(\phi^\dag_S\phi_S)_1(\phi^\dag_S\phi_S)_1 + \la^{\phi_S}_{2}(\phi^\dag_S\phi_S)_{1'}(\phi^\dag_S\phi_S)_{1''} \crn
& +  \la^{\phi_S}_{3}(\phi^\dag_S\phi_S)_{3_A}(\phi^\dag_S\phi_S)_{3_A}+\la^{\phi_S}_{4}(\phi^\dag_S\phi_S)_{3_S}(\phi^\dag_S\phi_S)_{3_S} \crn
&+ \la^{\phi_S}_{5}(\phi^\dag_S\phi_S)_{3_S}(\phi^\dag_S\phi_S)_{3_A},\crn
V(\phi_T,\phi_S)=&\la^{TS}_1(\phi^\dag_T\phi_T)_{1}(\phi^\dag_S\phi_S)_{1}+[\la^{TS}_2(\phi^\dag_T\phi_T)_{1'}(\phi^\dag_S\phi_S)_{1''}+\textrm{H.c.}] \crn
&+\la^{TS}_3(\phi^\dag_T\phi_T)_{3_A}(\phi^\dag_S\phi_S)_{3_A} +\la^{TS}_4(\phi^\dag_T\phi_T)_{3_S}(\phi^\dag_S\phi_S)_{3_S}\crn
&+\la^{TS}_5(\phi^\dag_T\phi_S)_{1}(\phi^\dag_S\phi_T)_{1}+\la^{TS}_6(\phi^\dag_T\phi_S)_{1'}(\phi^\dag_S\phi_T)_{1''}\crn
& +\la^{TS}_7(\phi^\dag_T\phi_S)_{3_A}(\phi^\dag_S\phi_T)_{3_A}+\la^{TS}_{8}(\phi^\dag_T\phi_S)_{3_S}(\phi^\dag_S\phi_T)_{3_S}\crn
&+[\la^{TS}_{9}(\phi^\dag_T\phi_S)_{3_A}(\phi^\dag_S\phi_T)_{3_S}+\textrm{H.c.}],\crn
V(\phi_T,h_u)&=\la^{Tu}(\phi^\dag_T\phi_T)_1(h_u^{\dag}h_u),\crn
V(\phi_T,h_d)&=\la^{Td}(\phi^\dag_T\phi_T)_1(h_d^{\dag}h_d),\crn
V(\phi_S,h_u)&=\la^{Su}(\phi^\dag_S\phi_S)_1(h_u^{\dag}h_u),\crn
V(\phi_S,h_d)&=\la^{Sd}(\phi^\dag_S\phi_S)_1(h_d^{\dag}h_d),\crn
V(\phi_T,\xi',\xi^{''})&=\la_1^{T\xi'\xi'}(\phi_T^{\dag}\phi_T)_{1}(\xi'^\ast\xi')+\la_2^{T\xi''\xi''}(\phi_T^{\dag}\phi_T)_{1}(\xi''^\ast\xi'')\crn& +[\la_3^{T\xi'\xi''}(\phi_T^{\dag}\phi_T)_{1''}(\xi'^\ast\xi'')_{1'}+\textrm{H.c.}],\crn
V(\phi_S,\xi',\xi^{''}) &= \la_1^{S\xi'\xi'}(\phi_S^{\dag}\phi_S)_{1}(\xi'^\ast\xi')+\la_2^{S\xi''\xi''}(\phi_S^{\dag}\phi_S)_{1}(\xi''^\ast\xi'')\crn
&+[\la_3^{S\xi'\xi''}(\phi_S^{\dag}\phi_S)_{1''}(\xi'^\ast\xi'')_{1'}+\textrm{H.c.}].
\label{Hpotential1}
\end{align}
There are ten neutral Higgs components in the model, implying ten equations for the minimal condition of the Higgs potential in Eq. (\ref{Hpotential}).  But only nine  equations are independent of each other, namely
\begin{align*}
&\mu_1^2 + \la^{u\xi'}u'^2 + \la^{u\xi''}u''^2 + \la_3v_d^2 + 3\la^{Su}v_S^2 + \la^{Tu}v_T^2 + 2\la_1v_u^2=0,\\
&\mu_2^2+\la^{d\xi'}u'^2 + \la^{d\xi''}u''^2 + 2\la_2v_d^2 + 3\la^{Sd}v_S^2 + \la^{Td}v_T^2 + \la_3v_u^2 = 0,\\
& \mu_3^2 u' + 2 \la^{\xi'}u'^3 + \la^{\xi'\xi''}u' u''^2 + \la^{d\xi'}u'v_d^2 + 3\la_1^{S\xi'\xi'}u'v_S^2 + 3\la_3^{S\xi'\xi''}u''v_S^2 \crn & +\ \la_1^{T\xi'\xi'}u'v_T^2+\la^{u\xi'}u' v_u^2=0,\\
& \mu_4^2 u'' + 2 \la^{\xi''}u''^3 + \la^{\xi'\xi''}u'^2 u'' + \la^{d\xi''}u''v_d^2 + 3\la_3^{S\xi'\xi''}u'v_S^2 + 3\la_2^{S\xi''\xi''}u''v_S^2 \crn & +\ \la_2^{T\xi''\xi''}u''v_T^2+\la^{u\xi''}u'' v_u^2=0,\\
& \mu_T^2+  \la_1^{T\xi'\xi'}u'^2 + \la_2^{T\xi''\xi''}u''^2 + \la^{Td}v_d^2 +3 \la_1^{ST}v_S^2 + \la_5^{ST}v_S^2 \crn & + \ \la_6^{ST}v_S^2 + 2\la_7^{ST} v_S^2+ 6\la_8^{ST}v_S^2 + 2\la_1^Tv_T^2 + 8\la_4^Tv_T^2 + \la^{Tu}v_u^2 =0,\\
& \la_3^{T\xi'\xi''}u'u'' +3 \la_2^{ST}v_S^2 + \la_5^{ST}v_S^2 +  \la_6^{ST}v_S^2- \la_7^{ST} v_S^2-3\la_8^{ST}v_S^2 =0,\\
& \mu_S^2+  \la_1^{S\xi'\xi'}u'^2 +\la_3^{S\xi'\xi''}u'u'' +\la_2^{S\xi''\xi''}u''^2 + \la^{Sd}v_d^2 +6 \la_1^{ST}v_S^2 +6 \la_2^{S}v_S^2 \crn &+\ \la_1^{ST}v_T^2+4 \la_4^{ST}v_T^2 + \ \la_5^{ST}v_T^2 + 4\la_8^{ST} v_T^2+ \la^{Su}v_u^2 =0,\\
& \mu_S^2+  \la_1^{S\xi'\xi'}u'^2 + 2\la_3^{S\xi'\xi''}u'u'' +\la_2^{S\xi''\xi''}u''^2 + \la^{Sd}v_d^2 + 6 \la_1^{S}v_S^2 + 6 \la_2^{S}v_S^2 \crn &+\ \la_1^{ST}v_T^2-2 \la_4^{ST}v_T^2 +  \la_6^{ST}v_T^2 + \la_7^{ST} v_T^2 + \la_8^{ST} v_T^2-2 \la_9^{ST} v_T^2 + \la^{Su}v_u^2 =0,\\
& \mu_S^2 +  \la_1^{S\xi'\xi'}u'^2 + 2\la_3^{S\xi'\xi''}u'u'' + \la_2^{S\xi''\xi''}u''^2 + \la^{Sd}v_d^2 +6 \la_1^{S}v_S^2 +6 \la_2^{S}v_S^2 \crn &+\ \la_1^{ST}v_T^2-2 \la_4^{ST}v_T^2 + \la_7^{ST} v_T^2+ \la_8^{ST} v_T^2+2 \la_9^{ST} v_T^2+ \la^{Su}v_u^2 =0.
\end{align*}
These correspond to nine dependent parameters which are represented as functions of the remaining parameters in the Higgs potential, including the VEVs of  neutral Higgs components. The nine dependent parameters chosen in this work are $\mu^2_1$, $\mu^2_2$, $\mu^2_3$, $\mu^2_4$, $\mu^2_T$, $\mu^2_S$, $\la_4^{ST}$, $\la_6^{ST}$, and $\la_3^{T\xi'\xi''}$. Inserting them into Eq. (\ref{Hpotential}), the Higgs potential contains only independent parameters. The assumed vacuum aligments given in Table~\ref{particle content} satisfy the above minimal equations,  hence this assumption can be dynamically achieved.
Now,  we can  find the masses and mass eigenstates of Higgs bosons predicted by the model.

Regarding CP-odd neutral Higgs components,  it is easily shown that the squared mass matrix  has a zero determinant, which implies  exactly a massless sate corresponding to the Goldstone boson of the $Z$ boson in the SM.
On the other hand, this model must contain at least one SM-like Higgs bosons observed by the LHC. Hence, the squared mass matrix of the CP-even neutral Higgs bosons must contain this Higgs boson. The squared mass matrix of the CP-even Higgs components is a $10\times 10$ matrix with a large number of  Higgs self-couplings which are independent parameters. In this work we will choose a simple case of the Higgs potential  that makes the Higgs spectrum realistic. In other words, the Higgs potential must satisfy the following conditions: (i) boundedness from below (BFB) and vacuum stability, (ii) all masses of physical Higgs are positive, (iii) having an SM-like Higgs boson observed by the LHC.  Here we will focus mainly on the identification of an SM-like Higgs boson.

In general, the squared mass matrix of the CP-even Higgs bosons are a $10\times 10$ matrix, where the main contribution to the SM-like Higgs boson arises from the two Higgs doublets $h_u$ and $h_d$. Hence, we will choose the regime  that these Higgs doublets decouple to other Higgs singlets, namely
\bea
\la^{u\xi'}= \la^{d\xi'}= \la^{u\xi''}= \la^{d\xi''},\hs
\la^{Tu}=\la^{Td} = \la^{Su} =
\la^{Sd}=0. \eea
With this choice, the mass matrix will separate into two submatrices, a $2\times2$ and an $8\times8$. The $8\times8$ matrix gives eight physical heavy Higgs bosons with masses depending on heavy VEVs $v_S$ and $v_T$. In the original basis $(S_u,\;S_d)^{\textrm{T}}$, the $2\times2$ matrix contains an SM-like Higgs boson and has the form
\bea
M^2_1=\left(\begin{array}{cc}
	4\la_1v_u^2 &  2 \la_3v_uv_d\\
	2 \la_3v_uv_d  &  4\la_2v_d^2 \\
\end{array}\right).
\eea
This gives two mass eigenstates, denoted as $H_1$ and $H_2$. Their masses and relations with the original states are
\begin{align}
& m^2_{H_1}=2 v^2c^2_{\beta}\left[ \la_1t_{\beta}^2+\la_2-\sqrt{\left(\lambda_1  t^2_{\beta}-\lambda_2\right)^2+\lambda_3^2t^2_{\beta}}\right], \; H_1=  S_uc_\alpha- S_ds_\alpha, \crn
%\label{smHiggs}\\
& m^2_{H_2}=2 v^2c^2_{\beta}\left[ \la_1t_{\beta}^2+\la_2+\sqrt{\left(\lambda_1  t^2_{\beta}-\lambda_2\right)^2+\lambda_3^2t^2_{\beta}}\right], \;  H_2=S_u s_\alpha + S_d c_\alpha,
\label{NHiggs}
\end{align}
where $ s_{\alpha}\equiv\sin\alpha$, $c_{\alpha}\equiv\cos\alpha$, $s_{2\alpha}\equiv\sin2\alpha$, and
\bea  \tan2\alpha=\dfrac{\la_3 t_\beta}{\la_2 -\la_1 t^2_\beta}.
\label{smhmix}\eea
In the limit $\beta=\alpha+\pi/2$, we can show that the couplings of $H_1$ with other SM particles are the same as the SM predictions. Hence, in our model $H_1$ is identified with the SM-like Higgs boson found experimentally.

Regarding CP-odd neutral Higgs components,  it is easily shown that the squared mass matrix  has a zero determinant, which implies  exactly a massless sate corresponding to the Goldstone boson of the $Z$ boson in the SM. Nine other CP-odd neutral Higgs are irrelevant to the phenomenology mentioned in this work.
%%%%%%%%%%%%%%%%%%%%%%%%%%%%%%%%%%%%%%%%%%%%%%%%%%%%%%%%%%%%%%%

\section{\label{CLR1}Passarino$-$Veltman functions for LFV decays $e_b\rightarrow e_a\gamma$ ($b>a$)}
The  Passarino$-$Veltman functions, called $C$-functions, are defined as follows:
\bea
C_{0,\mu,\mu\nu} (M_1,M_2,M_2) \equiv \frac{1}{i\pi^2}\int \frac{d^4
	k\times \{1,k_\mu,k_{\mu\nu}\}}{D_0D_1D_2},\label{oneloopin1}\eea
%-----
where $D_0=k^2-M_1^2,\;D_1=(k+p_b)^2-M_2^2$, and $D_2=(k+p_a)^2-M_2^2$, where $p_b\equiv\,p_1$ and $p_a\equiv p_2$ in usual notations for definitions of $C_{0,i,ij}$. The scalar $C$-functions are defined as $C_{\mu}=C_1p_{b\mu}+ C_2p_{a\mu}$ and $C_{\mu\nu}= C_{00} g_{\mu\nu} + C_{11}p_{b\mu}p_{b\nu}+ C_{12}(p_{b\mu}p_{a\nu}+ p_{b\nu}p_{a\mu})+ C_{22}p_{a\mu}p_{a\nu} $. For LFV decay processes $e_b\rightarrow e_a\gamma$ we denote $p_{a,b}^2=m_{a,b}^2$, where $m_{a,b}$ are the  masses of the charged leptons $e_{a,b}$. The  momentum of the photon $q=p_b-p_a$ satisfies $(p_b-p_a)^2=q^2=0$. The $C$-functions in this case are,
\bea
C_0&=& \frac{t-1-t\ln t}{M_2^2(t-1)^2}, \;
C_1=C_2=- \frac{3t^2-4t+ 1-2t^2\ln t}{4(t-1)^3M_2^2},\crn
C_{11}&=& C_{22}=2C_{12}=\frac{11t^3-18t^2+ 9t -2 -6t^3\ln t}{18M_2^2(t-1)^4},
\label{nCf}\eea
where $t=M_1^2/M_2^2$. With $t=1$, we have  $C_0=-1/(2M_2^2)$, $C_1=1/(6M_2^2)$ and $C_{11}=-1/(12M_2^2)$.

The definition of derivatives in Eq. (\ref{derivative}) results in the definition of the tensor strength of gauge bosons as  $F^a_{\mu\nu}=\partial_{\mu}W^a_{\nu}-\partial_{\nu}W^a_{\mu}+g \epsilon_{abc}W^b_{\mu}W^c_{\nu}$. The  couplings of the photon to $W^\pm$ and $\phi^\pm$ are then determined as follows:
% check 24/sep/2019
\begin{align}
	A_{\mu}\varphi^+\varphi^-: &\ ie(p_+-p_-)^{\mu} ,\crn
	A_{\lambda}W^+_{\mu}W^-_{\nu}:&\ -ie \left[g^{\lambda\mu}\left(q-p_+\right)^{\nu} +g^{\mu\nu}\left(p_+-p_-\right)^{\lambda}    + g^{\nu\lambda}\left(p_- -  q\right)^{\mu} \right]
	\label{eq_Gaugecoup}
\end{align}
where $q$ and $p_{\pm}$ denote incoming photon momenta  and $\varphi^{\pm}(W^{\pm})$, respectively.

Contributions from $W$ and $\varphi^\pm$ bosons to $C_{L,R}$  are calculated based on the general form given in Ref.~\cite{Hue:2017lak},
% a=3,b=2, check 24/sep/2019
\bea
C^{W}_{L} &=& -\frac{ e g^2m_a}{32\pi^2m_W^2}\sum_{i=1}^6U^{\nu}_{bi}U^{\nu*}_{ai}\left[ 2( C_{12} + C_{22} -C_1) m_W^2 + m_b^2(C_{11} + C_{12} + C_1) \right.\crn
&& \left. + m_{n_i}^2 (C_0 +C_1 +2 C_2+ C_{12} + C_{22} )\right], \crn
C^{W}_{R} &=& -\frac{ e g^2m_b}{32\pi^2 m_W^2}\sum_{i=1}^6U^{\nu}_{bi}U^{\nu*}_{ai}\left[2( C_{11} + C_{12}- C_2) m_W^2+ m_a^2 (C_{12} + C_{22} + C_2) \right.\crn
&& \left. + m_{n_i}^2(C_0 +  2C_1 +C_2+ C_{11} + C_{12} )\right]\eea
with $C_{0,a,ab}=C_{0,a,ab}(m_{n_i},m_W,m_W)$,
and
\bea
C^{\varphi}_{L} = -\frac{m_a e g^2}{32\pi^2m^2_W}
\sum_{i=1}^6U^{\nu*}_{ai}U^{L}_{bi}&\times&\left\{ t_{\beta}^{2}m^2_{b}(C_1 +C_{11}+C_{12})\right.\crn
&+&\left. m^2_{n_i}\left[t_{\beta}^{-2}\left(C_{2}+C_{12} +C_{22}\right) -(C_0 +C_1+C_2) \right]\right\}, \crn
C^{\varphi}_{R} =-\frac{m_b e g^2}{32\pi^2m^2_W}\sum_{i=1}^6U^{\nu*}_{ai}U^{L}_{bi}&\times&\left\{ t_{\beta}^{2}m^2_{a}(C_2 + C_{12}+C_{22})\right.\crn
&+&\left. m^2_{n_i}\left[t_{\beta}^{-2}\left(C_1 +C_{11}+C_{12}\right)  -(C_0 +C_1+C_2 )\right]\right\}\eea
with $C_{0,a,ab}=C_{0,a,ab}(m_{n_i},m_{\varphi},m_{\varphi})$.

The formula for $C^{W}_{L,R}$ is consistent with that given in Refs. \cite{He:2002pva, Ibarra:2011xn,Dinh:2012bp,Petcov:2013poa} in the limit $m_a,m_b << m_W$ and  $t_{W,i}\equiv \frac{m^2_{n_i}}{m^2_{W}}$, namely
\bea
\frac{C^W_L}{m_a}= \frac{C^W_R}{m_b}=- \frac{g^2e}{32\pi^2 m_W^2} f_V(t_{Wi}), \quad f_V(t)=-\frac{10- 43 t+78 t^2- 49t^3+4 t^4 +18 t^3\ln t }{12\left(t-1\right)^4}. \eea
In the limit $m_{a,b}\rightarrow 0$ with $t_{\beta}=1$ and $t_{\varphi,i}=\frac{m^2_{n_i}}{m^2_{\varphi}}$,  the  contributions from the charged Higgs bosons $C^{\varphi}_{L,R}$ have the following forms:
\bea
\frac{C^{\varphi}_L}{m_a}= \frac{C^{\varphi}_R}{m_b}=- \frac{g^2e}{32\pi^2 m_W^2} f_{s}(t_{\varphi,i}),\quad f_{s}(t) \equiv \frac{t\left[7 -12t -3 t^2 +8t^3 -6t\left( -2+3t \right)\ln t  \right] }{12 \left(t-1\right)^4}. \eea

%%%%%%%%%%%%%%%%%%%%%%%%%%%%%%%%%%%%%%%%%%%%

\end{document}